\documentclass[journal]{IEEEtran}

\usepackage[colorlinks=true,linkcolor=blue!50!black,citecolor=blue!40!black,urlcolor=blue!50!black]{hyperref}
\usepackage{graphicx}
\usepackage{amsmath,amssymb}
\usepackage{booktabs}
\usepackage{xcolor}
\usepackage{listings}
\usepackage{enumitem}
\usepackage{microtype}
\usepackage{cite}
\let\citep\cite
\let\citet\cite
\usepackage{algorithm}
\usepackage{algpseudocode}
\usepackage{dblfloatfix}   
\usepackage{orcidlink}

\definecolor{todoRed}{RGB}{180,30,30}

\newcommand{\localfig}{figures}

\title{Joint elastic full waveform inversion of multi-component geophone\\and distributed acoustic sensing data}

\author{Hoang Anh Nguyen\orcidlink{0000-0002-7504-5284} and Ali Tura%
\thanks{H. A. Nguyen is with the Department of Geophysics, Colorado School of Mines, Golden, CO 80401, USA (e-mail: hoanganh\_nguyen@mines.edu).}%
\thanks{A. Tura is with the Department of Petroleum Engineering, Colorado School of Mines, Golden, CO 80401, USA.}%
}

\begin{document}

\maketitle

\begin{abstract} Joint full waveform inversion (FWI) of distributed acoustic sensing (DAS) and ocean-bottom node (OBN) data typically requires converting measured strain to particle velocity, introducing numerical noise and spectral distortion. To eliminate this, we present an elastic multi-parameter FWI framework using a velocity--stress--strain (VSS) formulation that directly models pressure, particle velocity, and gauge-length-averaged DAS strain from a single forward simulation. Data residuals are injected additively into a single backward simulation, making computational cost independent of the active sensor subsets. We benchmark individual and combined datasets on cross-talk and elastic Marmousi models. Our results show that joint inversion recovers elastic parameters more accurately than single deployments when the sensors offer complementary information. Specifically, pairing two-component geophones with a deviated borehole DAS cable yields the most accurate parameter recovery and mitigates inter-parameter cross-talk by providing a distinct physical observable and complementary depth aperture. We release our implementation as \texttt{xFWI}, an open-source, Devito-based Python package for scalable, multi-deployment inversions.\end{abstract}

\begin{IEEEkeywords}
elastic FWI, DAS, multi-component, multi-parameter, joint inversion
\end{IEEEkeywords}

\section{Introduction}
\label{sec:intro}

Distributed acoustic sensing (DAS) has emerged as a mainstream geophysical recording technology \citep{Wang2018,Mateeva2014,Li2023} over the past decade. By utilizing permanently installed fiber-optic cables, DAS delivers dense, continuous strain measurements that effectively complement the coarser, discrete sampling of conventional point sensors. In modern offshore monitoring, particularly for CO$_2$ storage and oil and gas reservoirs, operators are increasingly combining multi-component ocean-bottom node (OBN) data with seabed DAS data. This integration enables a unified inversion framework, leveraging both data types simultaneously rather than processing them in isolation \citep{Liu2025}.

Hydrophones and multi-component geophones inside OBNs record pressure~$p$ and particle velocity~$\mathbf v$, respectively. DAS instead measures strain along an optical cable. It records the axial strain $\epsilon_{nn} = \hat n_i\,\hat n_j\,\epsilon_{ij}$, the strain tensor projected onto the local cable tangent $\hat{\mathbf n}$, averaged over a cable gauge length~$L$. All three datasets sense the same seismic wavefield, but in different physical quantities.

Full waveform inversion (FWI)~\citep{Virieux2009,Pratt1998} fits a forward-modelled wavefield to the recorded data and inverts the subsurface elastic parameters through a sequence of gradient-driven model updates. Multi-component geophone seismic data provide complementary sensitivity to different elastic parameters, helping to mitigate the inherent cross-talk in multi-parameter inversion~\citep{Brossier2009,Operto2013,Tura2025,Kohn2014,Modrak2016}. 

Joint inversion of DAS and multi-component geophone data has been pursued for near-surface imaging~\citep{Pan2019}, frequency-domain multi-parameter elastic recovery with shaped fiber-optic cables~\citep{Eaid2020}, and time-lapse monitoring~\citep{Lellouch2020}, establishing that DAS strain can be assimilated alongside geophone data to improve elastic recovery. Existing FWI workflows handle the heterogeneity by converting the recorded strain to particle velocity before computing the data residual. The strain-to-velocity conversion is either a wavenumber-domain integration with a singular kernel \citep{Sayed2020} or a numerical spatial differentiation; both introduce gauge-length-correction noise and spectral distortion that are particularly damaging at low signal-to-noise ratio. However, the conversion-free route, however, remains narrow: the direct-DAS inversion of \citet{Zhou2024} assimilates strain in its native quantity -- avoiding the strain-to-velocity conversion -- but was demonstrated for only a single strain component and a single parameter ($V_s$).  What is missing is a formulation that keeps this native-strain assimilation yet scales to the multi-parameter, multi-component setting that realistic monitoring acquisitions require.

In this work, a time-domain velocity--stress--strain (VSS) formulation casts pressure, particle velocity, and DAS strain as direct outputs of a single elastic forward simulation. Each quantity is modelled in the same form it is recorded and no strain-to-velocity conversion is required.  The residuals of all recorded data types are then injected additively into one single adjoint simulation. Therefore, any combination of deployments shares a single forward and a single backward solve, each added sensor type costing only a subdominant per-channel averaging rather than another wave-propagation solve.  The resulting inversion is multi-parameter $(V_p, V_s, \rho)$.

We evaluate the framework on controlled cross-talk and the elastic Marmousi benchmarks, inverting the same six recording subsets on both: hydrophone pressure $p$, multi-component geophones $[v_x, v_z]$, a seabed DAS cable co-located with the geophones ($\epsilon_{xx}$), a deviated borehole DAS cable ($\epsilon_{nn}$), and the joint combinations $[\epsilon_{xx}, \epsilon_{nn}]$ and $[v_x, v_z, \epsilon_{nn}]$.  Across all benchmarks, the joint geophone--borehole inversion recovers $(V_p, V_s, \rho)$ most accurately and most strongly suppresses inter-parameter cross-talk.

\section{Methods}
\label{sec:methods}

\subsection{Forward problem: VSS elastic wave equation}
\label{sec:forward}

We adopt a VSS formulation of the 2-D isotropic elastic wave equation.  The conventional velocity--stress system of \citet{Virieux1986} is retained as the primary forward engine, and the strain components $(e_x, e_z, e_s)$, with $e_x \equiv \epsilon_{xx}$, $e_z \equiv \epsilon_{zz}$, $e_s \equiv 2\,\epsilon_{xz}$, are tracked as a passive companion: strain is driven by velocity through the kinematic relation $\partial_t \epsilon_{ij} = \tfrac{1}{2}(\partial_i v_j + \partial_j v_i)$, but never feeds back into $\mathbf v$ or $\boldsymbol{\sigma}$.  The pure velocity--strain formulation \citep{Zhou2024} discards the stress fields and rewrites the momentum equations in terms of strains; we keep both.  This preserves the standard velocity--stress convolutional perfectly matched layer (CPML)~\citep{Komatitsch2007}; gives pressure as the stress trace rather than through the constitutive identity $-K(e_x + e_z)$, with $K$ the bulk modulus, that the strain-only form requires; and, critically for the discrete adjoint, keeps the density and stiffness coefficients in separate equations.  Velocity, stress, and strain are all advanced together in a single forward simulation:
\begin{equation}
\left\{\;
\begin{aligned}
  \rho\,\partial_t v_x &= \partial_x \sigma_{xx} + \partial_z \sigma_{xz} + f_x\,\delta_s, \\
  \rho\,\partial_t v_z &= \partial_x \sigma_{xz} + \partial_z \sigma_{zz} + f_z\,\delta_s, \\
  \partial_t \sigma_{xx} &= (\lambda + 2\mu)\,\partial_x v_x + \lambda\,\partial_z v_z + f_p\,\delta_s, \\
  \partial_t \sigma_{zz} &= \lambda\,\partial_x v_x + (\lambda + 2\mu)\,\partial_z v_z + f_p\,\delta_s, \\
  \partial_t \sigma_{xz} &= \mu\,(\partial_z v_x + \partial_x v_z), \\
  \partial_t e_x &= \partial_x v_x, \\
  \partial_t e_z &= \partial_z v_z, \\
  \partial_t e_s &= \partial_z v_x + \partial_x v_z,
\end{aligned}
\right.
\label{eq:vel-stress}
\end{equation}
where $\delta_s \equiv \delta(\mathbf x - \mathbf x_s)$ is the Dirac delta at source location $\mathbf x_s$, $\{f_x, f_z\}$ are the external body-force signatures, $f_p$ is the pressure-source signature, and $\lambda, \mu$ are the Lam\'e parameters.  The first five rows are the velocity--stress system; the last three are the kinematic strain evolution.  The factor-of-two difference between the strain-rate row $\partial_t e_s = \partial_z v_x + \partial_x v_z$ and the symmetric kinematic relation $\partial_t \epsilon_{xz} = \tfrac{1}{2}(\partial_z v_x + \partial_x v_z)$ is precisely the convention $e_s \equiv 2\epsilon_{xz}$ used throughout.

All recorded data streams used in this work (pressure, particle-velocity components, and the DAS strain channel) emerge from a single forward simulation, without a separate wave-propagation solve for each sensor type:
\begin{equation}
  p = -\tfrac{1}{2}(\sigma_{xx} + \sigma_{zz}),\qquad
  v_x,\;v_z,\;e_{nn}^{\rm DAS}\;\text{from the state.}
  \label{eq:data-from-state}
\end{equation}

The gauge-length-averaged DAS~\citep{Hartog2017} observable is built directly into the operator.  For a cable whose local tangent is the unit vector $\hat{\mathbf n} = (\hat n_x, \hat n_z)$ ($\hat n_x^2 + \hat n_z^2 = 1$) and gauge length $L$, define the axial-strain projection and its gauge-length average $\mathcal G_L$:
\begin{equation}
  \epsilon_{nn} \equiv \hat n_i\,\hat n_j\,\epsilon_{ij}
    = \hat n_x^2\, e_x + \hat n_z^2\, e_z + \hat n_x \hat n_z\, e_s,
  \label{eq:axial-strain}
\end{equation}
\begin{equation}
  e_{nn}^{\rm DAS}(\mathbf x_r, t) = \mathcal G_L(\epsilon_{nn}) \equiv
    \frac{1}{L}\!\int_{-L/2}^{L/2}\! \epsilon_{nn}(\mathbf x_r + \xi\,\hat{\mathbf n}, t)\,d\xi.
  \label{eq:gauge-length}
\end{equation}
The decomposition in \eqref{eq:axial-strain} uses the convention $e_s \equiv 2\epsilon_{xz}$ introduced above, so the off-diagonal coefficient is $\hat n_x \hat n_z$.  The average $\mathcal G_L$ is implemented as a boxcar window sum over the cells covered by the gauge length (an odd-length kernel for exact symmetry, so that $\mathcal G_L$ is self-adjoint, $\mathcal G_L^{\!\dagger} = \mathcal G_L$).  It is important to distinguish the strain field components $(e_x, e_z, e_s)$ of the forward state, which exist throughout the model, from the strain observable recorded by a DAS cable: a cable never measures a bare tensor component but the strain projected onto its local tangent $\hat{\mathbf n}$ and averaged over the gauge length $L$, i.e.\ the $e_{nn}^{\rm DAS}$ of \eqref{eq:gauge-length}. For a horizontal seabed cable ($\hat{\mathbf n} = \hat{\mathbf x}$, so $\hat n_x = 1$, $\hat n_z = 0$) this reduces to the gauge-length average of the $xx$ field component, $\epsilon_{xx}^{\rm DAS} \equiv \mathcal G_L(e_x)$ -- the projection of $e_x$ along $\hat{\mathbf x}$, not $e_x$ itself -- while the deviated borehole cable records the full tangential projection $\epsilon_{nn}^{\rm DAS}$.  Throughout the experiments, $\epsilon_{xx}$ and $\epsilon_{nn}$ denote these recorded DAS channels $\epsilon_{xx}^{\rm DAS}$ and $\epsilon_{nn}^{\rm DAS}$.

The forward, adjoint and gradient operators are implemented in Devito~\citep{Louboutin2019,Luporini2020}, which compiles the symbolic stencils to optimised CPU or GPU code at runtime.  The system is discretised on a staggered Yee-type grid: $(v_x,v_z)$ are staggered in their respective directions; $(e_x, e_z)$ live at the node; $e_s$ is staggered in both $x$ and $z$. Time integration is second-order leapfrog with time step $\Delta t$; spatial derivatives use the standard staggered finite-difference operator of order $2N$ ($N{=}4$ throughout this work, i.e.\ 8th-order spatial accuracy) with Holberg coefficients~\citep{Holberg1987}.  For an explicit leapfrog scheme on a staggered grid, stability requires the Courant--Friedrichs--Lewy (CFL) condition:
\begin{equation}
  V_p^{\max}\,\Delta t \le C^{(N)}\,\Delta x,
  \qquad
  C^{(N)} = \frac{1}{\sqrt{d}\,\sum_{m=1}^{N} |c_m|},
  \label{eq:cfl}
\end{equation}
where $V_p^{\max}$ is the maximum compressional velocity in the model, $d=2$ is the spatial dimension, and $\{c_m\}_{m=1}^{N}$ are the staggered finite-difference coefficients of half-order $N$ (the same staggered Holberg weights~\citep{Holberg1987} used in the spatial operator); the order-dependent Courant number $C^{(N)}$ is computed automatically from the operator coefficients and the model.  The same $C^{(N)}$ that fixes the stability limit here is reused to cap the upper $V_p$ box bound during inversion, so the forward stencil coefficients and the inversion stability ceiling are guaranteed to be consistent.

Energy reaching the absorbing boundary is attenuated by the CPML of \citet{Komatitsch2007}.  The complex coordinate stretch is implemented in the time domain through auxiliary memory-variable fields $\psi_j^{(\cdot)}$ that obey:
\begin{equation}
\left\{\;
\begin{aligned}
  \psi_j^{(\cdot),n+1} &= b_j\,\psi_j^{(\cdot),n} + a_j\,\partial_j(\cdot)^{n+1/2}, \\
  b_j &= e^{-(\sigma_j/\kappa_j + \alpha_j)\Delta t}, \\
  a_j &= \frac{\sigma_j(b_j-1)}{\sigma_j\kappa_j + \alpha_j\kappa_j^2},
\end{aligned}
\right.
\label{eq:cpml-update}
\end{equation}
with the standard CPML damping profile $\sigma_j(\xi) = \sigma_j^{\max}(\xi/L_{\rm pml})^p$, target reflection coefficient $R_c = 10^{-3}$, and frequency-shift parameter $\alpha_j$ that improves absorption at grazing incidence~\citep{Komatitsch2007}. The strain components require no separate absorbing treatment: being passive companions, their rates are built from the very same complex-stretched velocity gradients $(1/\kappa_j)\,\partial_j(\cdot) + \psi_j^{(\cdot)}$ and memory variables that update the stresses, so the recorded DAS strain inherits the CPML attenuation of the velocity--stress system directly.  This is a concrete benefit of retaining the velocity--stress engine rather than propagating strain as a primary field, for which the Komatitsch construction would have to be re-derived. The model stores the Lam\'e parameters $(\lambda,\mu,\rho)$, not the velocities~\eqref{eq:param-fwd}. This keeps the wave equation linear in the material state, which is what allows the same operator to serve as both forward and adjoint engine without re-derivation.

\subsection{Joint multi-component adjoint and gradient}
\label{sec:adjoint}

The joint misfit $J(\mathbf m)$ sums weighted $L_2$ residuals across the recorded data types $k \in \{p, v_x, v_z, e_{nn}^{\rm DAS}\}$:
\begin{equation}
  J(\mathbf m) = \sum_{k} \alpha_k \left[ \tfrac{1}{2}\!\sum_{s,r}\!\int_0^T\!\! |d_k^{\rm syn}(\mathbf m, t) - d_k^{\rm obs}(t)|^2\,dt \right],
  \label{eq:joint-misfit}
\end{equation}
each computed at the corresponding receiver locations from the forward state via \eqref{eq:data-from-state}--\eqref{eq:gauge-length}.  The DAS channel is the gauge-length average of the direction-cosine projection \eqref{eq:axial-strain}--\eqref{eq:gauge-length}, with $\hat n_x, \hat n_z$ the components of the cable tangent $\hat{\mathbf n}$.  The adjoint-state derivation below shows that the residuals enter a single backward simulation through additive injection rules.

Introduce adjoint multipliers $\mathbf v^\dagger = (v_x^\dagger, v_z^\dagger)$ for the momentum equations, $\boldsymbol{\sigma}^\dagger = (\sigma_{xx}^\dagger, \sigma_{zz}^\dagger, \sigma_{xz}^\dagger)$ for the stress evolution, and $\mathbf e^\dagger = (e_x^\dagger, e_z^\dagger, e_s^\dagger)$ for the kinematic strain tracker, all integrated backward in time on $[T, 0]$ with zero terminal conditions.  Stationarity of the augmented functional with respect to the forward state yields the adjoint system:
\begin{equation}
\left\{\;
\begin{aligned}
  \rho\,\partial_t v_x^\dagger &= \partial_x \sigma_{xx}^\dagger + \partial_z \sigma_{xz}^\dagger + \partial_x s_{ex}^\dagger \\
  &\qquad + \tfrac{1}{2}\partial_z s_{es}^\dagger + \textstyle\sum_r r_{v_x}\,\delta_r, \\
  \rho\,\partial_t v_z^\dagger &= \partial_x \sigma_{xz}^\dagger + \partial_z \sigma_{zz}^\dagger + \partial_z s_{ez}^\dagger \\
  &\qquad + \tfrac{1}{2}\partial_x s_{es}^\dagger + \textstyle\sum_r r_{v_z}\,\delta_r, \\
  \partial_t \sigma_{xx}^\dagger &= (\lambda+2\mu)\,\partial_x v_x^\dagger + \lambda\,\partial_z v_z^\dagger - \tfrac{1}{2}\textstyle\sum_r r_p\,\delta_r, \\
  \partial_t \sigma_{zz}^\dagger &= \lambda\,\partial_x v_x^\dagger + (\lambda+2\mu)\,\partial_z v_z^\dagger - \tfrac{1}{2}\textstyle\sum_r r_p\,\delta_r, \\
  \partial_t \sigma_{xz}^\dagger &= \mu\,(\partial_z v_x^\dagger + \partial_x v_z^\dagger),
\end{aligned}
\right.
\label{eq:adjoint}
\end{equation}
where the factor $\tfrac{1}{2}$ on the pressure residual is the Jacobian $\partial p/\partial\sigma_{xx} = \partial p/\partial\sigma_{zz} = -\tfrac{1}{2}$ of the pressure observable $p = -\tfrac{1}{2}(\sigma_{xx}+\sigma_{zz})$ in 2-D (it generalises to $1/d$ in $d$ dimensions, with $p \equiv -\tfrac{1}{d}\,\mathrm{tr}\,\boldsymbol{\sigma}$), $\delta_r \equiv \delta(\mathbf x - \mathbf x_r)$ and $\delta_d \equiv \delta(\mathbf x - \mathbf x_d)$ are the Dirac deltas at the point-receiver and DAS-channel locations, respectively, together with the strain-adjoint accumulators defined as fields:
\begin{equation}
\left\{\;
\begin{aligned}
  s_{ex}^\dagger(\mathbf x, t) &= \textstyle\sum_d \hat n_x^2\,\Delta e_{nn}\,\delta_d, \\
  s_{ez}^\dagger(\mathbf x, t) &= \textstyle\sum_d \hat n_z^2\,\Delta e_{nn}\,\delta_d, \\
  s_{es}^\dagger(\mathbf x, t) &= \textstyle\sum_d 2\,\hat n_x \hat n_z\,\Delta e_{nn}\,\delta_d,
\end{aligned}
\right.
\label{eq:adj-strain-accum}
\end{equation}
where $\Delta e_{nn} \equiv \alpha_{e_{nn}^{\rm DAS}}\,\mathcal G_L^{\!\dagger} \big(e_{nn}^{\rm syn} - e_{nn}^{\rm obs}\big)$ is the (gauge-length-spread, energy-weighted) recorded DAS residual for the channel at $\mathbf x_d$, and the spatial dimension is $d=2$.  The accumulator $s_{es}^\dagger$ is the adjoint source for the symmetric shear strain $\epsilon_{xz}$ (the field the forward operator evolves through $\partial_t \epsilon_{xz} = \tfrac{1}{2}(\partial_z v_x + \partial_x v_z)$), so its weight is $\partial e_{nn}^{\rm DAS}/\partial\epsilon_{xz} = 2\,\hat n_x \hat n_z$, the derivative of the projection $e_{nn}^{\rm DAS} = \mathcal G_L(\hat n_x^2 e_x + \hat n_z^2 e_z + 2\,\hat n_x \hat n_z\,\epsilon_{xz})$ with respect to $\epsilon_{xz}$; the two normal-strain weights are $\partial e_{nn}^{\rm DAS}/\partial e_x = \hat n_x^2$ and $\partial e_{nn}^{\rm DAS}/\partial e_z = \hat n_z^2$.  Equivalently, in the $e_s \equiv 2\epsilon_{xz}$ field of \eqref{eq:axial-strain} the off-diagonal Jacobian is $\partial e_{nn}^{\rm DAS}/\partial e_s = \hat n_x \hat n_z$; the factor of two between the two forms is exactly the $e_s = 2\epsilon_{xz}$ convention, and below it is paired with the $\tfrac{1}{2}$ that transposes the symmetric forward coupling.  These three weights map the single scalar residual $\Delta e_{nn}$ onto the three strain-adjoint accumulators that act as the point-source field at each channel.  The $(\mathbf v^\dagger, \boldsymbol{\sigma}^\dagger)$ subsystem mirrors the forward velocity--stress structure, so the same staggered-grid stencils are reused; the strain-adjoint accumulators carry the DAS residual to the velocity adjoint through the divergence terms $\partial_x s_{ex}^\dagger$, $\tfrac{1}{2}\partial_z s_{es}^\dagger$ (and their symmetric counterparts) in the first two rows of \eqref{eq:adjoint}.  These divergence terms are the discrete adjoint of the forward kinematic relation: differentiating $\partial_t e_x = \partial_x v_x$ and $\partial_t \epsilon_{xz} = \tfrac{1}{2}(\partial_z v_x + \partial_x v_z)$ with respect to $v_x$ transposes $\partial_x$ and $\partial_z$ into the source terms $\partial_x s_{ex}^\dagger$ and $\tfrac{1}{2}\partial_z s_{es}^\dagger$ that drive $v_x^\dagger$ (symmetrically for $v_z^\dagger$); the $\tfrac{1}{2}$ is the transpose of the $\tfrac{1}{2}$ in the symmetric forward shear-strain rate.  Both the $\tfrac{1}{2}$ coupling in \eqref{eq:adjoint} and the $2\,\hat n_x \hat n_z$ accumulator weight in \eqref{eq:adj-strain-accum} match the implementation, in which the shear-strain accumulator is stored for the tensor strain $\epsilon_{xz}$; the forward and adjoint coefficients are therefore mutually consistent.  The accumulators $s_{ex}^\dagger, s_{ez}^\dagger, s_{es}^\dagger$ are continuous fields, not bare distributions: each channel residual is first point-spread by $\delta_d$ to its location and gauge-spread over the $L$-cell window by $\mathcal G_L^{\!\dagger}$ inside $\Delta e_{nn}$, so the spatial derivatives $\partial_x s_{ex}^\dagger$ and $\partial_z s_{es}^\dagger$ in \eqref{eq:adjoint} act on these smooth, gauge-length-supported fields rather than differentiating a $\delta$-function directly.
Note that the operator is not self-adjoint when DAS is included: the forward strain tracker is passive (no feedback into $\mathbf v$ or $\boldsymbol{\sigma}$), but the adjoint of the kinematic relation reverses this direction, so the strain-adjoint accumulators must feed back into the velocity adjoint.  Without this back-coupling the DAS residual would never reach the model parameters, since the gradient \eqref{eq:grad} depends on correlations of forward fields with $\mathbf v^\dagger$ and $\boldsymbol{\sigma}^\dagger$ rather than with the strain accumulators directly.

The three point-receiver residuals (pressure and the two velocity components) enter through:
\begin{equation}
\left\{\;
\begin{aligned}
  r_p      &= \alpha_p\,(p^{\rm syn} - p^{\rm obs}), \\
  r_{v_x}  &= \alpha_{v_x}\,(v_x^{\rm syn} - v_x^{\rm obs}), \\
  r_{v_z}  &= \alpha_{v_z}\,(v_z^{\rm syn} - v_z^{\rm obs}),
\end{aligned}
\right.
\label{eq:five-injection}
\end{equation}
while the fourth data type, the DAS strain residual, enters through the strain-adjoint accumulators \eqref{eq:adj-strain-accum}.  Together \eqref{eq:five-injection} and \eqref{eq:adj-strain-accum} constitute the additive injection rules for all four recorded data types $\{p, v_x, v_z, e_{nn}^{\rm DAS}\}$ in the single backward simulation: any subset of deployments is selected simply by activating the corresponding residual terms.  In \eqref{eq:adj-strain-accum}, $\mathcal G_L^{\!\dagger}$ is the adjoint of the gauge-length averaging operator $\mathcal G_L$ in \eqref{eq:gauge-length}, a uniform spread of the residual over the $L$-cell window centred on the DAS channel; $\mathcal G_L^{\!\dagger} = \mathcal G_L$ (self-adjoint, as in \eqref{eq:gauge-length}).

The weights $\alpha_k$ equalise initial-residual energy across data types so that no single deployment dominates the early-stage gradient:
\begin{equation}
  \alpha_k = \big[\textstyle\sum_{s,r}\!\int_0^T |d_k^{\rm syn,(0)} - d_k^{\rm obs}|^2\,dt\big]^{-1}.
  \label{eq:alpha-equalise}
\end{equation}

The Lam\'e-space gradient is the time-correlation between forward and adjoint wavefields~\citep{Tarantola1984,Plessix2006}:
\begin{equation}
\left\{\;
\begin{aligned}
  \partial_\lambda J &= \int_0^T (\partial_x v_x + \partial_z v_z)\,(\sigma_{xx}^\dagger + \sigma_{zz}^\dagger)\,dt, \\
  \partial_\mu     J &= \int_0^T\!\big[\,2\,\partial_x v_x\,\sigma_{xx}^\dagger + 2\,\partial_z v_z\,\sigma_{zz}^\dagger \\
                     &\qquad\quad + (\partial_z v_x + \partial_x v_z)\,\sigma_{xz}^\dagger\big]\,dt, \\
  \partial_\rho    J &= \int_0^T\!\big(v_x\,\partial_t v_x^\dagger + v_z\,\partial_t v_z^\dagger\big)\,dt.
\end{aligned}
\right.
\label{eq:grad}
\end{equation}
The chain-rule projections from $(\lambda, \mu, \rho)$ to $(V_p, V_s, \rho)$ used in the experiments are derived in the next subsection.  A single forward simulation computes \eqref{eq:vel-stress}, a structurally identical adjoint simulation runs in reverse time with the injection rule \eqref{eq:five-injection}, and the gradient expressions are evaluated as time-correlation imaging conditions during the adjoint simulation.  Because all sensor types share the one forward and the one backward solve, enabling or disabling a data type changes only the receiver-side bookkeeping (which residuals are formed and injected, plus the gauge-length averaging for DAS channels) and leaves the dominant wave-propagation cost unchanged; the per-iteration cost is therefore essentially independent of which subset of components is included.

\subsection{Parameterizations}
\label{sec:param}

The forward operator is written in Lam\'e parameters, so $\nabla_{\!\mathbf m_{\!\lambda}} J$ with $\mathbf m_{\!\lambda} = (\lambda, \mu, \rho)$ is the primary gradient returned by the adjoint simulation.  Inversion in velocity $\mathbf m_{\!v} = (V_p, V_s, \rho)$ or in acoustic--shear impedance $\mathbf m_{\!I} = (I_p, I_s, \rho)$ therefore needs two maps: (i) $\mathbf m_{\!v}\mapsto\mathbf m_{\!\lambda}$ for the forward run, and (ii) its Jacobian-transpose $(\partial\mathbf m_{\!\lambda}/\partial\mathbf m_{\!v})^{\!\top}:\nabla_{\!\mathbf m_{\!\lambda}}J\mapsto\nabla_{\!\mathbf m_{\!v}}J$ for the gradient.

The forward maps are:
\begin{equation}
\left\{\;
\begin{aligned}
  \lambda &= (V_p^2 - 2 V_s^2)\,\rho, \\
  \mu     &= V_s^2\,\rho,
\end{aligned}
\right.
\qquad
\left\{\;
\begin{aligned}
  \lambda &= (I_p^2 - 2 I_s^2)/\rho, \\
  \mu     &= I_s^2/\rho.
\end{aligned}
\right.
\label{eq:param-fwd}
\end{equation}
Differentiating these gives the chain-rule projections of the Lam\'e-space gradient into velocity space:
\begin{equation}
\left\{\;
\begin{aligned}
  \partial_{V_p}    J &= 2\rho V_p\,\partial_\lambda J, \\
  \partial_{V_s}    J &= -4\rho V_s\,\partial_\lambda J + 2\rho V_s\,\partial_\mu J, \\
  \partial_{\rho}^{(v)} J &= (V_p^2 - 2 V_s^2)\,\partial_\lambda J + V_s^2\,\partial_\mu J + \partial_\rho J,
\end{aligned}
\right.
\label{eq:grad-vel}
\end{equation}
and into impedance space:
\begin{equation}
\left\{\;
\begin{aligned}
  \partial_{I_p}    J &= \tfrac{2 I_p}{\rho}\,\partial_\lambda J, \\
  \partial_{I_s}    J &= -\tfrac{4 I_s}{\rho}\,\partial_\lambda J + \tfrac{2 I_s}{\rho}\,\partial_\mu J, \\
  \partial_{\rho}^{(I)} J &= -\tfrac{I_p^2 - 2 I_s^2}{\rho^2}\,\partial_\lambda J
                             - \tfrac{I_s^2}{\rho^2}\,\partial_\mu J + \partial_\rho J.
\end{aligned}
\right.
\label{eq:grad-imp}
\end{equation}
Two structural consequences follow:  first, the velocity-space $V_s$ gradient pulls from $\partial_\lambda J$ and $\partial_\mu J$ with opposite signs; this is the algebraic origin of the $V_p$--$V_s$ cross-talk that the radiation pattern alone cannot explain; second, the impedance density gradient subtracts the $(I_p^2 - 2 I_s^2)/\rho^2$ part of $\partial_\lambda J$ out of $\partial_\rho^{(I)} J$, the algebraic content of Tarantola's classical observation that impedance decouples density~\citep{Tarantola1986}.  In our experiments the inversion is run in the velocity parameterization \eqref{eq:grad-vel} so that the $V_p$ and $V_s$ physical constraints and the CFL ceiling reduce to native box bounds on $V_p$ and $V_s$.

\subsection{Inversion}
\label{sec:inversion}

The model is inverted in $(V_p, V_s, \rho)$ space with a multiscale schedule~\citep{Bunks1995}: at each stage the observed data and the source wavelet are bandpass-filtered to a low--high corner pair $(f_\ell^{(i)}, f_h^{(i)})$ and a bound-constrained quasi-Newton optimiser~\citep{Byrd1995} drives the model toward a stationary point of $J$ before the next band is added.  Per-shot gradients are stacked after an energy-based diagonal preconditioner~\citep{Plessix2006} normalises the geometric-spreading and illumination-amplitude bias:
\begin{equation}
\begin{aligned}
  \mathbf g &\leftarrow \text{diag}\big(P(\mathbf x) + \varepsilon\big)^{-1}\,\mathbf g, \\
  P(\mathbf x) &= \Big[\textstyle\sum_s W_s(\mathbf x)\Big]^{1/2}\!
                \int_{\Gamma_r}\!\frac{ds_r}{|\mathbf x - \mathbf x_r|},
\end{aligned}
  \label{eq:precond}
\end{equation}
\begin{equation}
  W_s(\mathbf x) = \int_0^T |\mathbf v_s(\mathbf x, t)|^2\,dt,
  \label{eq:precond-energy}
\end{equation}
where $\sum_s W_s(\mathbf x)$ is the time-integrated forward-wavefield energy summed over shots and $\int_{\Gamma_r} |\mathbf x - \mathbf x_r|^{-1}\,ds_r$ is the receiver-side geometric-spreading factor obtained from the analytic line integral of the Green's-function amplitude over the receiver array $\Gamma_r$, following \citet{Plessix2004}; their product is a pseudo-Hessian diagonal that approximates the combined source- and receiver-side geometric-spreading and illumination-amplitude bias, and $\varepsilon$ is a small stabiliser that prevents division by zero in poorly illuminated cells.  This receiver-Green's-function form is the preconditioner used in all runs reported here; for the curved borehole cable the receiver integral is evaluated as a discrete sum of $|\mathbf x - \mathbf x_r|^{-1}$ over the channel positions along the cable trajectory.  A spatial taper masks the perfectly matched layer (PML) halo and a small radius around each source/receiver to suppress the singular punch-through artefact at injection sites.

Three physical requirements are imposed on the velocity model at every iterate of the bound-constrained optimiser.  First, simple box bounds keep each parameter in a physically admissible range, $V_p \in [V_p^{\min}, V_p^{\max}]$, $V_s \in [V_s^{\min}, V_s^{\max}]$, and $\rho \in [\rho^{\min}, \rho^{\max}]$.  Second, the upper $V_p$ bound is capped at the CFL ceiling \eqref{eq:cfl}, $V_p^{\max} \le C^{(N)}\,\Delta x / \Delta t$, so no update can push the model past the explicit-time-stepping stability limit of the forward solver.  Third, bulk-modulus positivity is enforced as a coupled constraint on the $V_p/V_s$ ratio rather than as an independent box bound: $K = \lambda + \tfrac{2}{3}\mu \ge 0$ is equivalent to $(V_p/V_s)^2 \ge 4/3$, and at the start of each stage the iterate is projected onto:
\begin{equation}
  V_s \le V_p \big/ r, \qquad r = (1 + \eta)\,\tfrac{2}{\sqrt 3},
  \label{eq:k-positivity}
\end{equation}
where $2/\sqrt 3 \approx 1.155$ is the physical $K = 0$ ratio and a small safety margin $\eta = 0.04$ ($r \approx 1.20$) holds the solution off the cliff where the bulk modulus vanishes and the elastic operator degenerates.  This protected threshold, not the larger ratio $V_p/V_s = \sqrt 2$ at which $\lambda = (V_p^2 - 2V_s^2)\rho$ changes sign, is the relevant one: a negative $\lambda$ is physically admissible for $2/\sqrt 3 < V_p/V_s < \sqrt 2$ (auxetic-like behaviour with $K > 0$), so it is $K \ge 0$, not $\lambda \ge 0$, that guards operator non-degeneracy and stability.  Without equation \eqref{eq:k-positivity}, an unconstrained $V_s$ update can drive $V_p/V_s$ below $2/\sqrt 3$, making the bulk modulus negative and the forward solve unstable.  Finally, because L-BFGS-B applies a single step length across the entire model vector, each parameter block is normalised by its maximum magnitude before concatenation into the optimiser variable, so the disparate $(V_p, V_s, \rho)$ magnitudes do not bias the search direction~\citep{Byrd1995}.

\subsection{Distributed computing on a cluster}
\label{sec:distributed}

The per-shot forward and gradient evaluations are independent and together dominate the FWI run time, so the campaign is run as a coordinator--worker task graph.  A single coordinator process holds the current model $\mathbf m$ and the optimiser state; for every objective evaluation it scatters the per-shot work over $N_{\rm w}$ worker nodes that each run one forward--adjoint pair locally and return the per-shot gradient and misfit.  The coordinator stacks the gradients, hands the result to the optimiser, receives the update, and broadcasts the new model to the workers for the next evaluation.  Algorithm~\ref{alg:dist} captures the data flow.

The implementation maps Algorithm~\ref{alg:dist} onto a Dask~\citep{Dask2016} task graph: workers are launched as a dynamically scaled cluster on the available job queue (e.g., using SLURM~\citep{10.1007/10968987_3}), the per-shot forward and adjoint routine runs inside a Dask task on each worker, and the coordinator process gathers the per-shot futures and stacks the gradients before passing the sum to the optimiser.  The number of workers is set at submission time and adjusted by the queue manager during the run, so the same launcher scales from a single CPU node to several hundred GPU workers with no source-code changes.  Operator just-in-time (JIT) compilation runs at most once per model shape per worker process, so the compilation cost is paid once at worker startup and amortised across every iteration of the multiscale schedule.

\begin{algorithm}
\caption{Shot-parallel FWI on a cluster}
\label{alg:dist}
\begin{algorithmic}[1]
\State \textbf{Coordinator} initialises $\mathbf m_0$, multiscale
       schedule $\{(f_\ell^{(i)}, f_h^{(i)})\}_{i=1}^{N_s}$, and
       spawns $N_{\rm w}$ workers.
\For{stage $i = 1, \ldots, N_s$}
  \State Bandpass-filter observed data and source wavelet to
         $[f_\ell^{(i)}, f_h^{(i)})$.
  \While{optimiser not converged}
    \State \textit{Coordinator}: broadcast $\mathbf m$ to all workers.
    \For{shot $s = 1, \ldots, N_{\rm shots}$ \textbf{in parallel}}
      \State \textit{Worker}: forward simulation \eqref{eq:vel-stress}
             with shot $s$; record synthetics.
      \State \textit{Worker}: residuals \eqref{eq:five-injection};
             adjoint simulation; per-shot gradient $\mathbf g_s$ and
             misfit $J_s$.
      \State \textit{Worker}: return $(\mathbf g_s, J_s)$ to coordinator.
    \EndFor
    \State \textit{Coordinator}: stack
           $\mathbf g \leftarrow \sum_s \mathbf g_s$,
           $\;J \leftarrow \sum_s J_s$.
    \State \textit{Coordinator}: apply preconditioner and taper;
           project to chosen parameterization (\S\ref{sec:param}).
    \State \textit{Coordinator}: bound-constrained quasi-Newton step;
           update $\mathbf m$.
  \EndWhile
\EndFor
\State \Return final $\mathbf m$.
\end{algorithmic}
\end{algorithm}

\section{Numerical experiments}
\label{sec:results}

We present numerical experiments to evaluate the performance of the different inversion approaches on synthetic seismic data. We first consider a controlled cross-talk benchmark, followed by a more realistic Marmousi model benchmark. We characterize the joint inversion through two benchmarks using the same forward operator and the same inversion procedure; only the active component subset of the misfit \eqref{eq:joint-misfit} changes between configurations.  No noise is added to the synthetic records.  Both benchmarks compare the same six recording subsets -- hydrophone pressure $p$, multi-component geophones $[v_x, v_z]$, a horizontal seabed DAS cable $\epsilon_{xx}$, a borehole DAS cable $\epsilon_{nn}$, and the two joint subsets all-DAS $[\epsilon_{xx}, \epsilon_{nn}]$ and geophone--borehole $[v_x, v_z, \epsilon_{nn}]$.

\subsection{Cross-talk benchmark}
\label{sec:results-crosstalk}

\begin{figure*}[tbp]
  \centering
  \includegraphics[width=\linewidth]{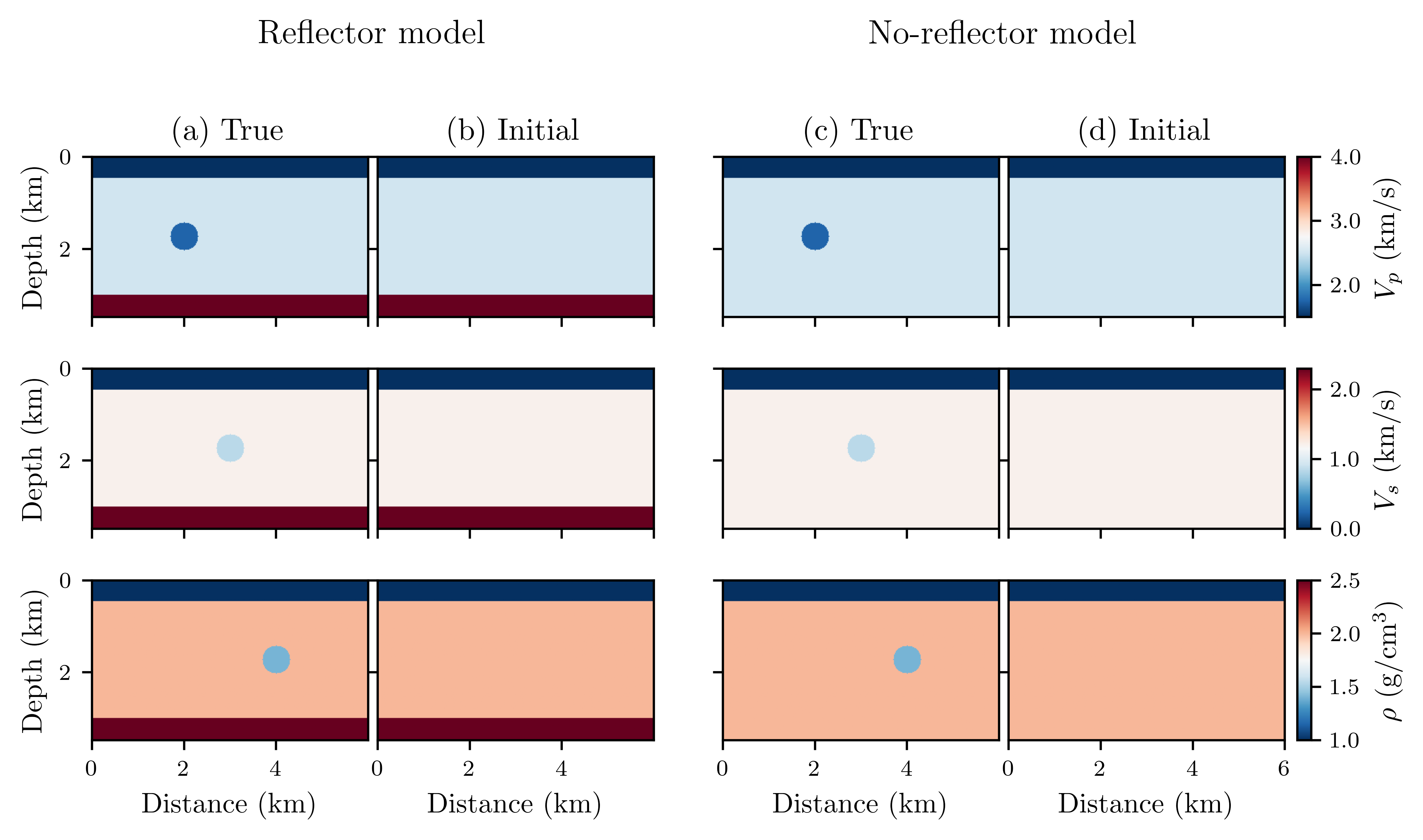}
  \caption{Cross-talk benchmark true and initial models for $V_p, V_s, \rho$
    (rows), for both variants used in this study: the with-reflector model
    (a)~true and (b)~initial -- inverted in
    Fig.~\ref{fig:crosstalk-platforms} and Table~\ref{tab:crosstalk-ssim} --
    and the no-reflector model (c)~true and (d)~initial -- inverted in
    Fig.~\ref{fig:crosstalk-borehole} and
    Table~\ref{tab:crosstalk-ssim}.  Each true model perturbs a single
    elastic parameter from the background ($V_p$ on the left, $V_s$ in the
    centre, $\rho$ on the right); the initial models contain the background
    layering only (no anomaly).  All panels share the with-reflector colour
    scale, so the deep high-$V_p$ half-space (the reflector, dark-red band) is
    present in (a, b) and absent in (c, d).}
  \label{fig:crosstalk-models}
\end{figure*}

\begin{figure}[tbp]
  \centering
  \includegraphics[width=\linewidth]{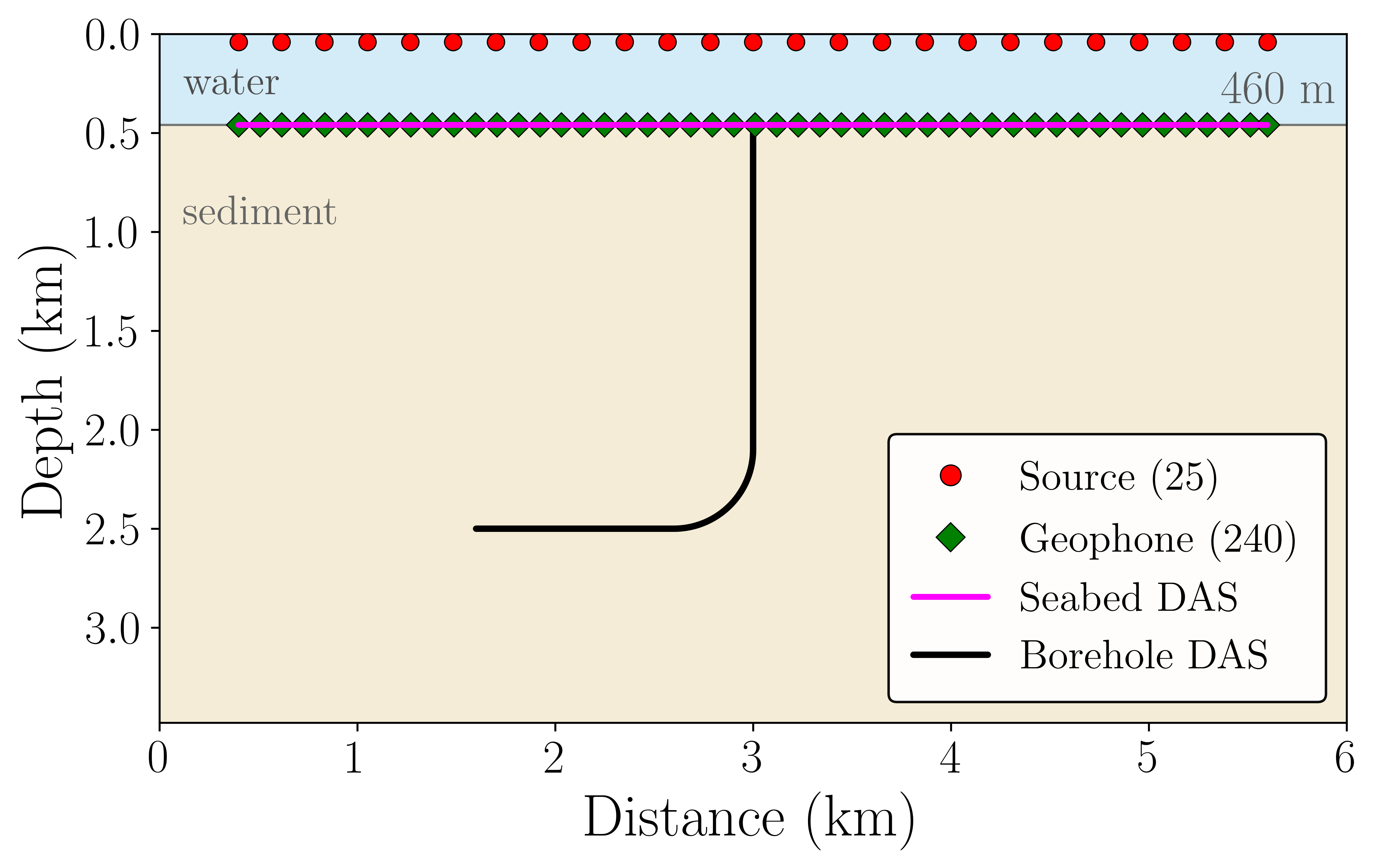}
  \caption{Cross-talk acquisition geometry (schematic): a water layer over the
    sediment.  25 airgun sources (red circles, $z = 40$~m), 240 geophone
    receivers (green diamonds, $z = 460$~m; every 5th drawn), the seabed DAS
    cable (magenta, co-located with the OBN array), and the borehole DAS cable
    (black) -- an L-shaped deviated cable that descends at $x = 3$~km and turns
    to a horizontal lateral at 2.5~km depth.}
  \label{fig:crosstalk-acq}
\end{figure}

In this experiment, we investigate a $6~\mathrm{km}\times 3.5~\mathrm{km}$ domain with a grid spacing of $\Delta x = \Delta z = 20$~m. The true and initial models are shown in Fig.~\ref{fig:crosstalk-models} with a water depth (bathymetry) of 460~m. The true model features a single circular anomaly in all elastic parameters. The inversions begin from an initial model that contains only the background layering and lacks the anomaly information. The acquisition geometry (Fig.~\ref{fig:crosstalk-acq}) consists of 25 explosive sources located in the water column at a depth of 40~m, along with 240 OBNs and both seabed and borehole DAS cables. Both DAS cables have a gauge length of $L = 10$~m. We apply a multiscale inversion strategy using frequency bands of 2, 5, 10, and 20~Hz, running for 100, 100, 50, and 50 iterations per stage, respectively. All six data subsets are inverted independently and in parallel using the velocity parameterization $(V_p, V_s, \rho)$. We use this parameterization throughout because it better suppresses inter-parameter cross-talk compared with the Lam\'e $(\lambda, \mu, \rho)$ parameterization~\citep{Kohn2014}.

\begin{figure*}[tbp]
  \centering
  \includegraphics[width=\linewidth]{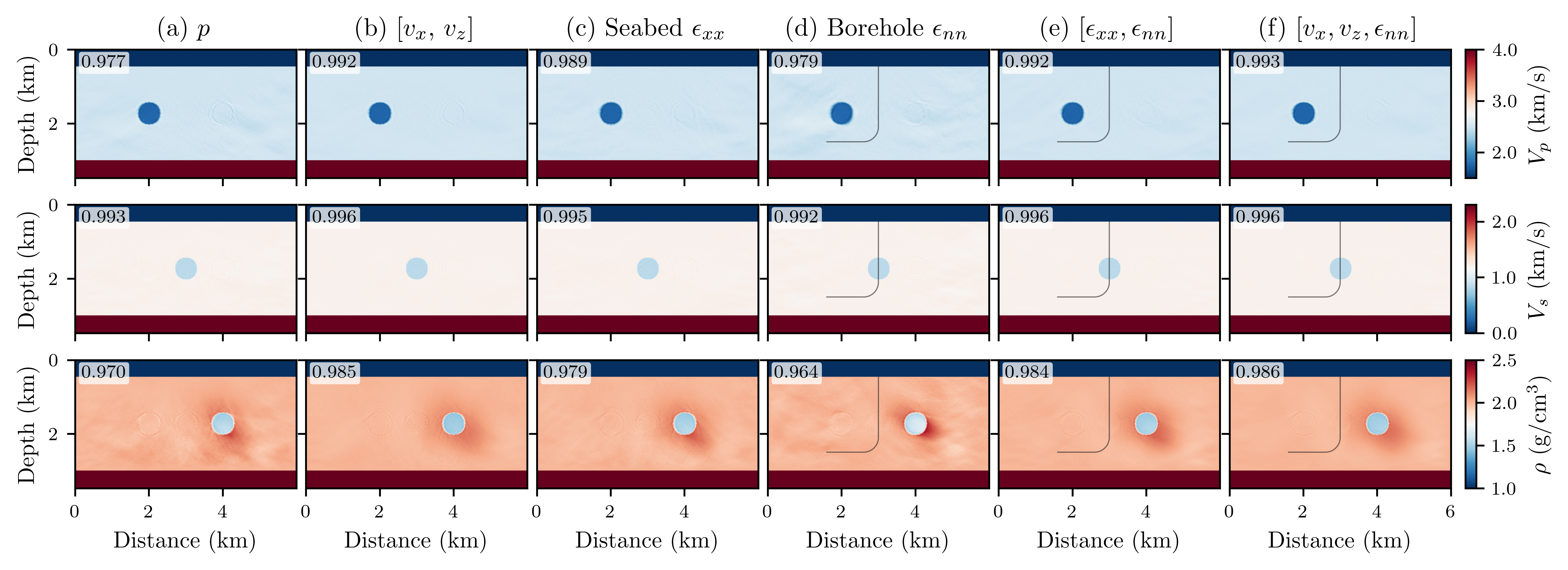}
  \caption{Inverted $V_p, V_s, \rho$ (rows) recovered at 20~Hz with the deep
    reflector retained, from (a)~pressure $p$, (b)~multi-component $[v_x, v_z]$, (c)~seabed
    DAS $\epsilon_{xx}$, (d)~borehole DAS axial strain $\epsilon_{nn}$,
    (e)~joint all-DAS $[\epsilon_{xx}, \epsilon_{nn}]$, and (f)~joint
    geophone--borehole $[v_x, v_z, \epsilon_{nn}]$.  The faint line in (d)--(f)
    traces the L-shaped borehole cable.  Per-panel numbers are the SSIM against
    the true model within the receiver aperture
    (Table~\ref{tab:crosstalk-ssim}).}
  \label{fig:crosstalk-platforms}
\end{figure*}

Fig.~\ref{fig:crosstalk-platforms} shows the 20~Hz FWI recovery of all six subsets with the deep reflector retained.  The half-space below the targets is stiffer and denser than the overlying sediment -- a jump in $V_p$, $V_s$ and $\rho$ at $z = 3$~km -- and its impedance contrast reflects most of the down-going energy back through the section.  Every deployment is therefore well illuminated and recovers the anomalies, the pressure-only inversion (a) least accurately and the two joint subsets (e, f) most accurately.

Table~\ref{tab:crosstalk-ssim} (top) reports the per-parameter Structural Similarity Index (SSIM) \citep{Wang2004} between the 20~Hz recovery and the true model. Because the strong deep reflector provides excellent illumination, every data subset achieves a high SSIM ($\geq 0.96$). Even the single-component recordings perform comparably to the two-component geophones: the seabed DAS cable results are nearly identical, while the pressure and borehole DAS data trail only marginally (most noticeably for density $\rho$). Although the joint geophone--borehole inversion remains nominally the most accurate across all parameters ($0.993, 0.996,$ and $0.986$ for $V_p, V_s,$ and $\rho$, respectively), its advantage over the best single component is very slight. For example, it ties with both the geophones and the seabed DAS cable for $V_s$ at $0.996$, and it outperforms the geophones for $V_p$ by only $0.001$. Ultimately, the strong reflection masks the differences in aperture and sensor type, meaning these minor third-decimal differences should be interpreted as near-ties rather than significant improvements.

\begin{figure*}[tbp]
  \centering
  \includegraphics[width=\linewidth]{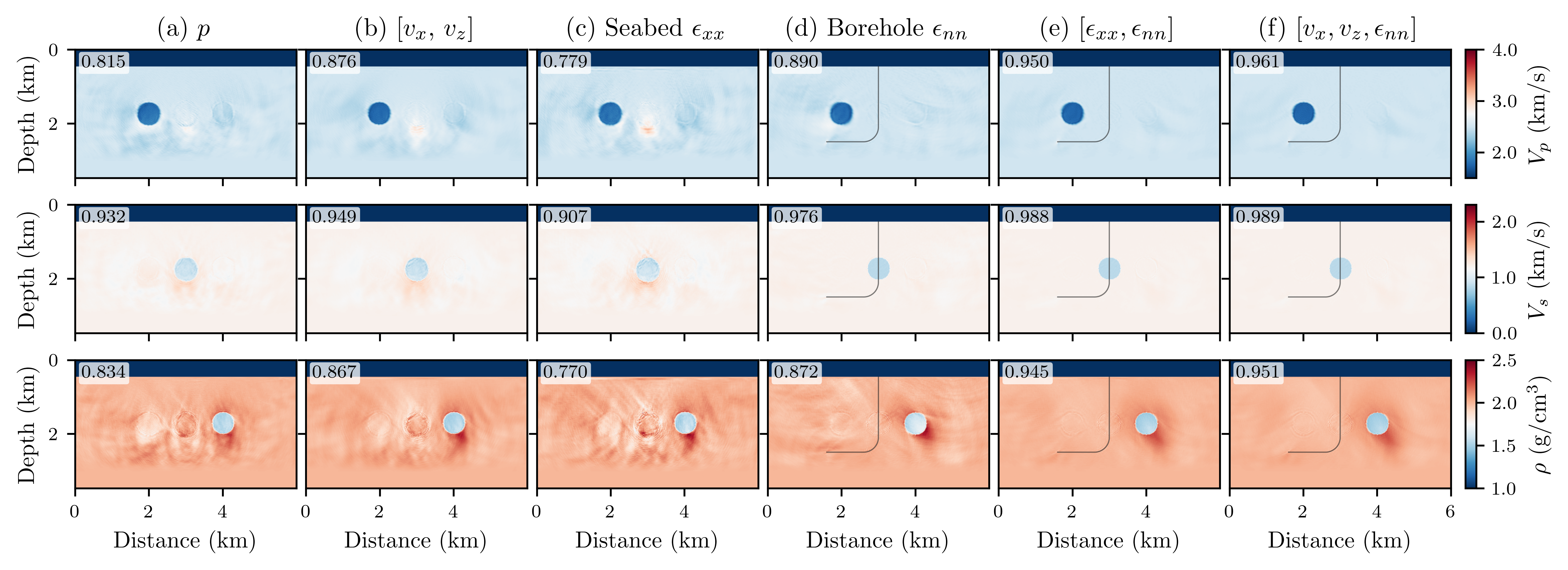}
  \caption{Inverted $V_p, V_s, \rho$ (rows) recovered at 20~Hz with the deep
    half-space removed, from (a)~pressure $p$, (b)~multi-component $[v_x, v_z]$, (c)~seabed
    DAS $\epsilon_{xx}$, (d)~borehole DAS axial strain $\epsilon_{nn}$,
    (e)~joint all-DAS $[\epsilon_{xx}, \epsilon_{nn}]$, and (f)~joint
    geophone--borehole $[v_x, v_z, \epsilon_{nn}]$.  The faint line in
    (d)--(f) traces the L-shaped borehole cable.  Per-panel numbers are the
    SSIM against the true model within the receiver aperture
    (Table~\ref{tab:crosstalk-ssim}).  Unlike the co-located case
    (Fig.~\ref{fig:crosstalk-platforms}), the differing borehole aperture lets
    the joint subsets exceed every single-deployment inversion.}
  \label{fig:crosstalk-borehole}
\end{figure*}

Removing the half-space eliminates this bottom-up illumination: the targets are now illuminated by the direct and diving wavefields alone, making the performance differences between deployments much more pronounced (Fig.~\ref{fig:crosstalk-borehole} and Table~\ref{tab:crosstalk-ssim}, bottom). Because its vertical-then-lateral aperture samples ray paths that the seabed sensors miss, the borehole DAS cable alone recovers the section more accurately than any individual seabed deployment. For instance, it outperforms the geophones, achieving SSIM values of $0.890, 0.976,$ and $0.872$ (for $V_p, V_s,$ and $\rho$, respectively) compared to the geophones' $0.876, 0.949,$ and $0.867$. The joint geophone--borehole subset is once again the most accurate across all parameters ($0.961, 0.989,$ and $0.951$), but this time by a substantial margin. This highlights a consistent mechanism that also governs the Marmousi results below: adding borehole data improves the recovery only marginally when a strong deep reflector already illuminates the section, but provides substantial uplift when that reflection is absent.

\begin{table}
  \centering
  \caption{SSIM between the 20~Hz inverted and true models for the cross-talk
    deployment/aperture comparison, by recording subset and elastic parameter,
    for both benchmarks: the with-reflector model
    (Fig.~\ref{fig:crosstalk-platforms}) and the no-reflector model
    (Fig.~\ref{fig:crosstalk-borehole}).  $\epsilon_{xx}$ is the
    seabed DAS cable, $\epsilon_{nn}$ the borehole DAS cable;
    higher is better and the highest value in each row is in bold.
    The SSIM support is a single mask held fixed across all six subsets
    within each benchmark, so the columns are mutually comparable; SSIM is a
    single structural scalar, and several with-reflector margins between
    subsets are at the third decimal.}
  \label{tab:crosstalk-ssim}
  \footnotesize\setlength{\tabcolsep}{3pt}
  \begin{tabular}{lcccccc}
    \toprule
    Parameter & $p$ & $[v_x, v_z]$ & $\epsilon_{xx}$ & $\epsilon_{nn}$ & $[\epsilon_{xx}, \epsilon_{nn}]$ & $[v_x, v_z, \epsilon_{nn}]$ \\
    \midrule
    \multicolumn{7}{l}{\emph{With reflector} (Fig.~\ref{fig:crosstalk-platforms})} \\
    $V_p$     & 0.977 & 0.992 & 0.989 & 0.979 & 0.992 & \textbf{0.993} \\
    $V_s$     & 0.993 & \textbf{0.996} & 0.995 & 0.992 & \textbf{0.996} & \textbf{0.996} \\
    $\rho$    & 0.970 & 0.985 & 0.979 & 0.964 & 0.984 & \textbf{0.986} \\
    \midrule
    \multicolumn{7}{l}{\emph{No reflector} (Fig.~\ref{fig:crosstalk-borehole})} \\
    $V_p$     & 0.815 & 0.876 & 0.779 & 0.890 & 0.950 & \textbf{0.961} \\
    $V_s$     & 0.932 & 0.949 & 0.907 & 0.976 & 0.988 & \textbf{0.989} \\
    $\rho$    & 0.834 & 0.867 & 0.770 & 0.872 & 0.945 & \textbf{0.951} \\
    \bottomrule
  \end{tabular}
\end{table}

\subsection{Marmousi benchmark}
\label{sec:results-marmousi}

\begin{figure*}[tbp]
  \centering
  \includegraphics[width=0.72\linewidth]{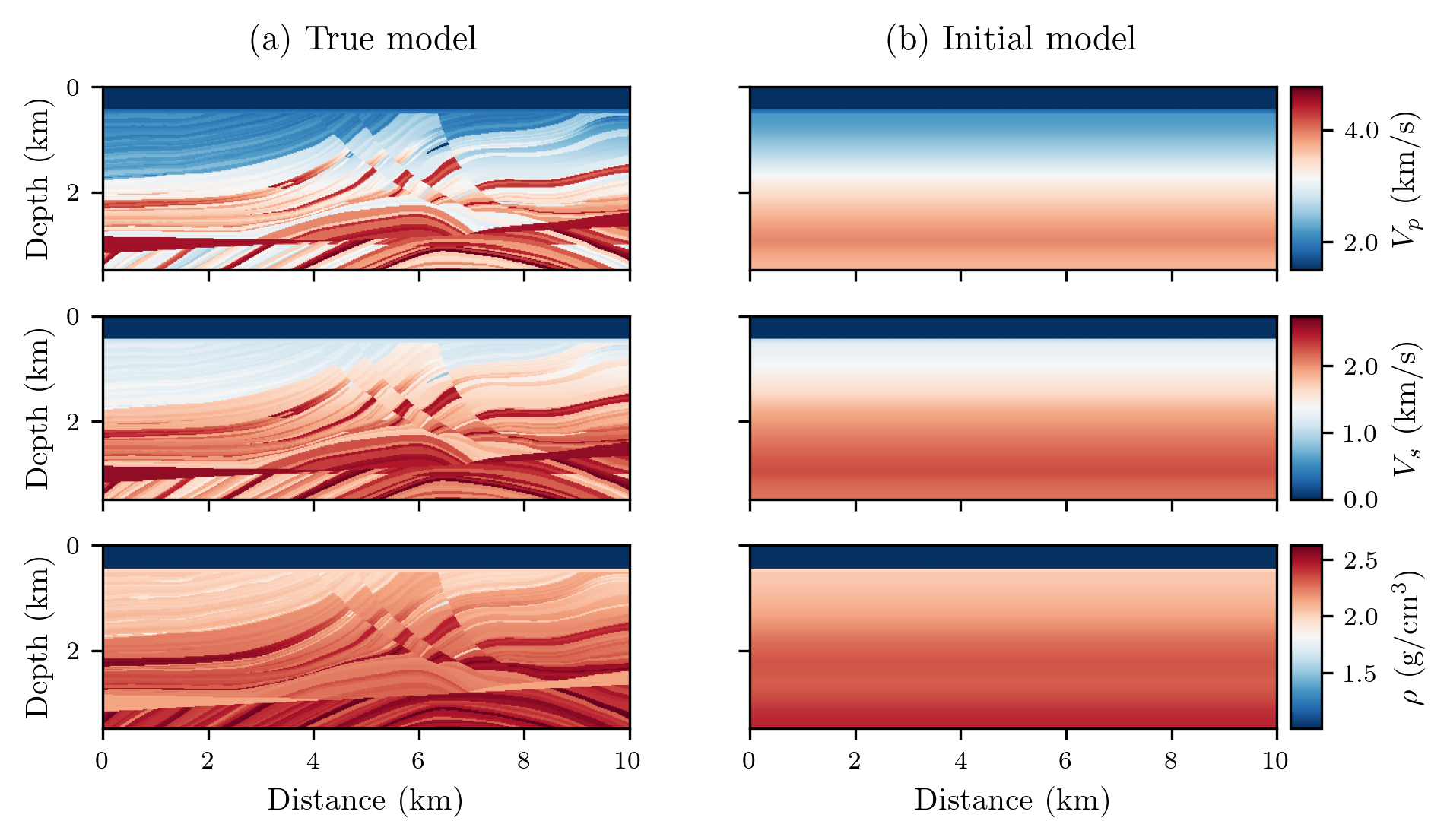}
  \caption{Marmousi (a) true and (b) initial models for
    $V_p, V_s, \rho$ (top to bottom).  The initial model is a smooth,
    laterally homogeneous $1$-D model that increases with depth (a
    depth gradient), identical for all inversions.}
  \label{fig:marmousi-trueinit}
\end{figure*}

We now extend this deployment comparison to a more realistic scenario: the elastic Marmousi model \citep{Martin2006} (Fig.~\ref{fig:marmousi-trueinit}), which has a grid spacing of $10$~m ($\Delta x = \Delta z = 10$~m). Every inversion in this section shares the same initial model, the same 50 near-surface airgun sources (distributed between $x = 0.8$ and $8.7$~km in the water column at a depth of $z = 40$~m), the same multiscale schedule (2, 5, 10, and 20~Hz, running for 100, 100, 50, and 50 iterations per stage, respectively), and the same inversion procedure. Only the recorded data subset changes between configurations.

\begin{figure*}[tbp]
  \centering
  \includegraphics[width=0.6\linewidth]{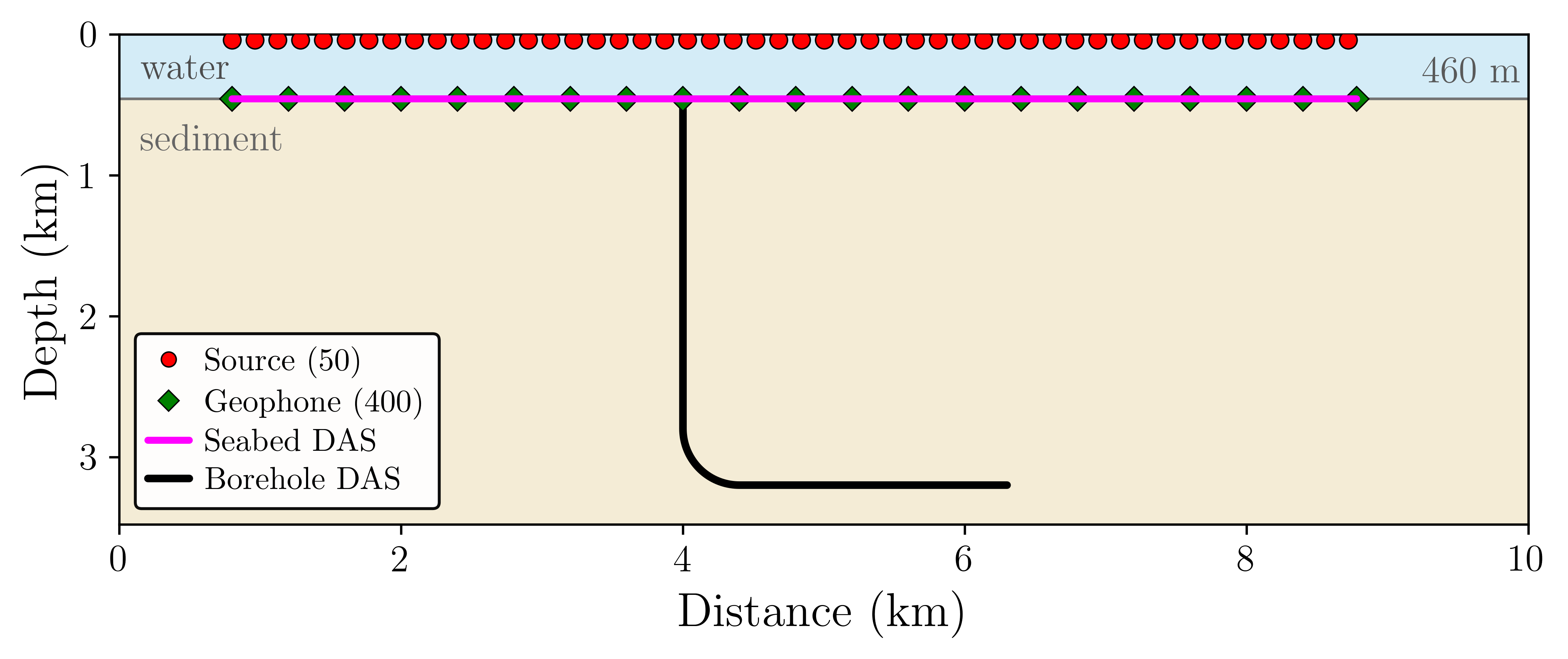}
  \caption{Marmousi acquisition geometry (schematic): a water layer over the
    sediment.  50 airgun sources (red circles, $z = 40$~m), 400 geophone
    receivers (green diamonds, $z = 460$~m; every 20th drawn), the seabed DAS
    cable (magenta, co-located with the OBN array), and the borehole DAS
    cable (black) -- an L-shaped deviated cable that descends at $x = 4$~km and
    turns to a horizontal lateral at $3.2$~km depth.}
  \label{fig:marmousi-acq}
\end{figure*}

The acquisition geometry (Fig.~\ref{fig:marmousi-acq}) mirrors the cross-talk benchmark. A seabed array at a depth of 460~m ($z = 460$~m) carries 400 OBNs spaced 20~m apart, spanning from $x = 0.8$ to $8.8$~km. Co-located with these OBNs is a horizontal seabed DAS cable recording $\epsilon_{xx}$ across 7981 channels (1~m spacing, $L = 10$~m gauge length). A separate borehole DAS cable follows an L-shaped deviated well: it descends vertically from the water bottom at $x = 4$~km down to a kickoff depth of $2.8$~km, bends through a $400$~m radius, and runs laterally at a depth of $3.2$~km out to $x = 6.3$~km, recording the axial strain $\epsilon_{nn}$ along its tangent. Because of this layout, the seabed DAS cable shares the same aperture as the geophones, whereas the borehole cable provides complementary sampling from deeper in the section. As before, the same six recording subsets defined in Section~\ref{sec:results} are inverted in the velocity parameterization.

\begin{figure*}[tbp]
  \centering
  \includegraphics[width=\linewidth]{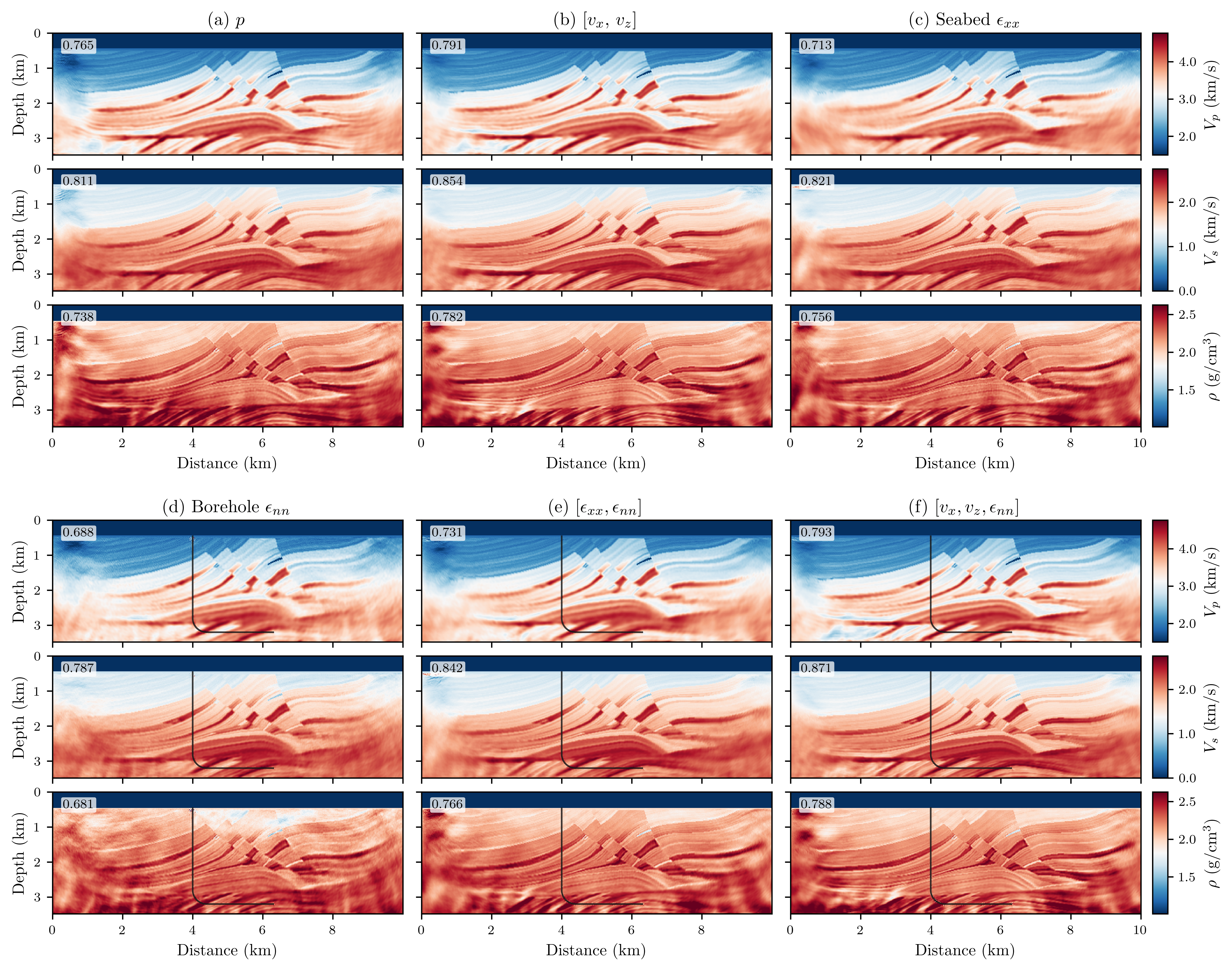}
  \caption{Inverted $V_p, V_s, \rho$ (rows) recovered at $20$~Hz on the
    Marmousi benchmark, from (a)~pressure $p$, (b)~multi-component $[v_x, v_z]$, (c)~seabed
    DAS $\epsilon_{xx}$, (d)~borehole DAS axial strain $\epsilon_{nn}$,
    (e)~joint all-DAS $[\epsilon_{xx}, \epsilon_{nn}]$, and (f)~joint
    geophone--borehole $[v_x, v_z, \epsilon_{nn}]$.  The faint line in
    (d)--(f) traces the L-shaped borehole cable.  Per-panel numbers are
    the SSIM against the true model within the receiver aperture
    (Table~\ref{tab:marmousi-ssim}).}
  \label{fig:marmousi-platforms}
\end{figure*}

Fig.~\ref{fig:marmousi-platforms} shows the $20$~Hz recovery for all six subsets, and Table~\ref{tab:marmousi-ssim} reports the corresponding per-parameter SSIM against the true model. The ranking echoes the cross-talk benchmark. Among the single deployments, the multi-component geophones are the strongest single choice, achieving SSIM values of 0.791, 0.854, and 0.782 for $V_p, V_s,$ and $\rho$, respectively. The seabed DAS cable trails the geophones across every parameter (yielding 0.713, 0.821, and 0.756), while the pressure data is even weaker for $V_s$ and $\rho$. Because the seabed cable and OBN deployments share the same aperture, the seabed cable provides little independent horizontal constraint; in fact, even the joint seabed-borehole DAS inversion fails to reach the multi-component geophone baseline.

\begin{figure*}[tbp]
  \centering
  \includegraphics[width=\linewidth]{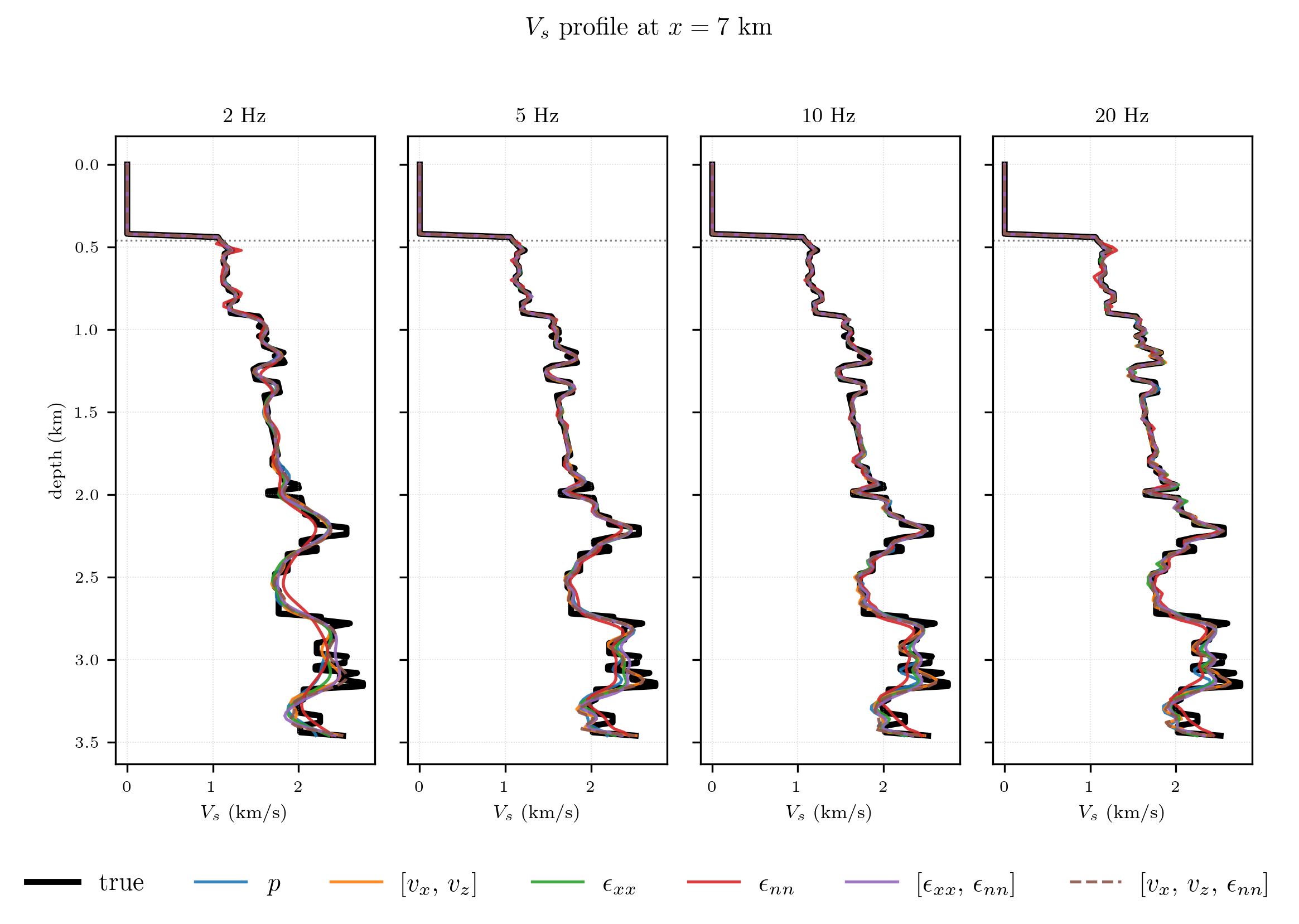}
  \caption{Vertical $V_s$ profiles at $x = 7$~km through the Marmousi
    model: the true model (thick black) and the six recovered subsets
    ($p$, $[v_x, v_z]$, $\epsilon_{xx}$, $\epsilon_{nn}$,
    $[\epsilon_{xx}, \epsilon_{nn}]$, $[v_x, v_z, \epsilon_{nn}]$) at
    each multiscale stage (2, 5, 10 and 20~Hz).  Successive stages add
    resolution, and by 20~Hz every subset reproduces the smooth
    background; the largest inter-subset differences appear in the
    deeper, more weakly illuminated section below $\sim$2.5~km.  The
    dotted line marks the seabed receiver depth (460~m).}
  \label{fig:marmousi-vs-profile}
\end{figure*}

The borehole cable supplies this missing complementary aperture. While the borehole record alone is the weakest single deployment, yielding SSIM values of 0.688, 0.787, and 0.681 (for $V_p, V_s,$ and $\rho$, respectively), this is expected because its narrow, localised aperture illuminates only a wedge of the section, leaving it unable to recover the full model as completely as the seabed deployments. However, when combined with the geophones, the deeper borehole aperture samples ray paths that the surface sensors miss. As a result, the joint geophone--borehole inversion attains the highest SSIM across all three parameters (0.793, 0.871, and 0.788), exceeding the multi-component baseline in every case. This improvement is decisive for $V_s$ (0.871 versus 0.854) and for the cross-talk-prone density $\rho$ (0.788 versus 0.782). For $V_p$, however, the joint margin over the geophone baseline is minimal (0.793 versus 0.791) and should be interpreted as a tie rather than a clear gain. As seen in the cross-talk benchmark, the complementary aperture of the borehole allows the joint inversion to surpass any single deployment, most notably in areas where geophone illumination is weak. Vertical $V_s$ profiles at $x = 7$~km (Fig.~\ref{fig:marmousi-vs-profile}) illustrate how each subset approaches the true model as the multiscale band widens, showing that the deployments differ most significantly in the deeper, weakly illuminated portions of the section.

\begin{table}[tbp]
  \centering
  \caption{SSIM between the $20$~Hz Marmousi recovery and the true
    model for each recording subset, computed within the receiver
    aperture.  $\epsilon_{xx}$ is the seabed DAS cable,
    $\epsilon_{nn}$ the borehole DAS cable; higher is
    better and the highest value in each row is in bold.  The joint
    geophone--borehole subset $[v_x, v_z, \epsilon_{nn}]$ is the
    highest in every parameter (see text for the per-parameter margins).  The
    SSIM support is a single mask held
    fixed across all six subsets, so the columns are mutually comparable;
    SSIM is a single structural scalar.}
  \label{tab:marmousi-ssim}
  \footnotesize\setlength{\tabcolsep}{3pt}
  \begin{tabular}{lcccccc}
    \toprule
    Parameter & $p$ & $[v_x, v_z]$ & $\epsilon_{xx}$ & $\epsilon_{nn}$ & $[\epsilon_{xx}, \epsilon_{nn}]$ & $[v_x, v_z, \epsilon_{nn}]$ \\
    \midrule
    $V_p$  & 0.765 & 0.791 & 0.713 & 0.688 & 0.731 & \textbf{0.793} \\
    $V_s$  & 0.811 & 0.854 & 0.821 & 0.787 & 0.842 & \textbf{0.871} \\
    $\rho$ & 0.738 & 0.782 & 0.756 & 0.681 & 0.766 & \textbf{0.788} \\
    \bottomrule
  \end{tabular}
\end{table}

\section{Discussion}
\label{sec:discussion}

\subsection{Physical drivers of parameter recovery}
\label{sec:disc-physics}

Each ground-motion observable primarily constrains the direction of motion it records. While a single horizontal ($v_x$) or vertical ($v_z$) channel captures only one projection of that motion, recording both components samples the full particle-motion vector. This physical completeness explains why the two-component geophone data consistently proves to be the strongest seabed subset across every parameter (Tables~\ref{tab:crosstalk-ssim} and~\ref{tab:marmousi-ssim}). The benchmarks also separate the data types by their exposure to cross-talk. For example, pressure data recovers $V_s$ and $\rho$ less accurately than multi-component data (yielding a Marmousi $\rho$ SSIM of 0.738 for pressure versus 0.782 for the geophones). This is expected because pressure predominantly carries acoustic P-wave energy---making it highly susceptible to $V_p$--$\rho$ cross-talk---whereas the two-component geophones also capture critical S-wave information.

Across all models and data subsets, the three elastic parameters rank consistently: $V_s$ is always the best recovered, while $\rho$ is the worst. The superior recovery of $V_s$ is a direct consequence of resolution: because shear velocity is lower than compressional velocity ($V_s < V_p$), the shear wavelength ($\lambda_s = V_s/f$) is correspondingly shorter than the compressional wavelength ($\lambda_p = V_p/f$). For a shared frequency band, the S-sensitive wavefield---constrained directly by the recorded shear motion---can therefore resolve structure at a much finer scale. Conversely, density is the most difficult parameter to recover. Its scattering radiation pattern overlaps strongly with both $V_p$ and $V_s$, and its overall perturbation of the wavefield is relatively weak. This leaves $\rho$ as the most cross-talk-exposed and least-constrained parameter, perfectly explaining its consistently trailing SSIM values.

\subsection{Co-located OBN and DAS deployments}
\label{sec:disc-redundancy}

A horizontal cable co-located on the OBN array measures axial strain $\epsilon_{xx} = \partial_x u_x$, which is the along-line spatial derivative of the horizontal displacement. A horizontal geophone inside the OBN, on the other hand, records the particle velocity $v_x = \partial_t u_x$. Because spatial differentiation inherently up-weights high wavenumbers, the two measurements are not exact algebraic identities. In principle, this means a co-located DAS cable could add independent, short-wavelength detail to the geophone data.

In this work, however, the OBN array already samples the wavefield very finely relative to the seismic wavelength. As a result, the cable's nominally finer spatial sampling captures no new arrivals that the OBN sensors miss. Empirically, we see that the seabed cable adds little information along the shared OBN aperture: as a standalone deployment, it never surpasses the two-component geophones (Tables~\ref{tab:crosstalk-ssim} and~\ref{tab:marmousi-ssim}), and in the strongly reflecting Marmousi model, even the all-DAS joint inversion fails to reach the geophone baseline. For this reason, we chose to pair the OBNs with a borehole cable rather than a second seabed cable.

It is important to note that this redundancy is driven not only by the dense OBN sampling used in our experiments, but also by the highly reflective geology of the Marmousi model. Because this multilayered geology generates abundant reflections, the OBNs are already richly illuminated and highly accurate on their own. In a real-world field survey, however, OBNs are often spaced much more coarsely. In such cases, the wavefield is no longer finely sampled along the array, meaning a co-located cable's dense sampling could finally realise that theoretical short-wavelength advantage. Furthermore, strict co-location is an artificial test condition; in field practice, a DAS line rarely follows the OBN spread exactly. Instead, it occupies its own trajectory, naturally sampling an aperture that the OBNs do not. The matched, co-located geometry analysed here is therefore a deliberately conservative case designed to suppress the very spatial complementarity from which joint inversion typically benefits.

\subsection{Unlocking joint gain through spatial complementarity}
\label{sec:disc-complementarity}

The benefit of joint inversion emerges once a deployment contributes information that is complementary to the seabed array. The borehole experiment clearly demonstrates this: while the OBNs and the seabed cable share a single array, adding the deviated borehole DAS cable makes the joint geophone--borehole inversion the strongest configuration across every parameter in both benchmarks. The magnitude of this gain tracks closely with how strongly reflections already illuminate the section. Where reflections are strong, the multi-component geophones are already highly accurate, meaning the borehole adds only modest value. Conversely, where reflections are weak, the seabed array is starved of information, and the borehole's complementary aperture becomes decisive. Because Marmousi's reflective, high-impedance geology falls into the strongly illuminated category, its joint gain is real but modest, with the $V_p$ recovery resulting in a tie (Table~\ref{tab:marmousi-ssim}).

Two distinct mechanisms can contribute to this gain. Geometrically, the borehole spans an aperture that the seabed sensors do not, illuminating reflection and diving-wave paths that neither deployment records on its own. Physically, strain and particle velocity project the $(V_p, V_s, \rho)$ scattering kernels onto different radiation patterns \citep{Tarantola1986}. By combining them, the inversion can constrain the density $\rho$ better than either observable can alone. Relative to the seabed array, introducing the borehole simultaneously changes the physical aperture, the recorded observable, and the ray-path coverage. In constrast, the co-located seabed cable changes the observable but not the aperture. It does not help with dense OBN. But when the OBN sparse compared to the seismic wavelength, it will help. Relative to the seabed array, introducing the borehole simultaneously changes the physical aperture, the recorded observables, and the ray-path coverage. In contrast, a co-located seabed cable changes the observables but not the aperture. Consequently, the cable provides little additional benefit when deployed alongside a dense OBN array. However, when the OBN spacing is sparse relative to the seismic wavelength, the inclusion of the seabed cable becomes highly advantageous.

\section{Conclusions}
\label{sec:conclusions}

We have presented a joint elastic multi-parameter FWI framework that seamlessly combines multi-component particle velocity and DAS strain without relying on data conversion or scaling approximations. Because the formulation natively evolves the full wavefield state, all recorded data streams emerge directly from the simulation, effectively eliminating the noisy strain-to-velocity conversions required by conventional DAS workflows. A single backward simulation efficiently handles every recorded data subset simultaneously through additive injection rules, making the per-iteration computational cost practically independent of which sensors are active. The framework also remains highly flexible, readily supporting individual inversions of hydrophone, geophone, or DAS data alone.

Carrying out the inversion in the velocity parameterization aligns the physical admissibility limits and the explicit time-stepping stability ceilings perfectly with the structure of our bound-constrained solver. Both constraints reduce to native bounds on $V_p$ and $V_s$ that hold exactly at every iteration, leaving only bulk-modulus positivity as a coupled constraint. In contrast, using the Lam\'e parameterization causes these limits to become coupled and non-linear, preventing them from being imposed as simple bounds. Ultimately, the velocity parameterization proves to be the natural choice for elastic FWI: it is fundamentally better conditioned against inter-parameter cross-talk, and it renders both stability and positivity limits enforceable as native box constraints.

Our experiments reveal that joint inversion recovers every elastic parameter more accurately than any single deployment, provided the combined sensors offer complementary information. Among the seabed deployments, the two-component geophones consistently prove to be the strongest single choice. Because a co-located seabed DAS cable shares this same aperture, it adds very little independent constraint. For this reason, we chose to pair the geophones with a borehole DAS cable rather than a second seabed cable. This joint geophone--borehole configuration achieved the best overall results. Relative to the seabed array, the borehole differs in both the aperture it samples and the physical observable it records. Because our experiments varied these two factors together, we attribute the inversion gains to the combination of a complementary depth aperture and a distinct strain observable. Disentangling the two would require an aperture-matched control experiment. Crucially, the fact that the co-located seabed cable changed the observable but added little value indicates that a distinct observable alone is insufficient under the dense, matched sampling conditions used here.

\section*{Computer code availability}
\label{sec:code}

All experiments in this paper were carried out using \texttt{xFWI}, an open-source Python implementation of the VSS formulation and the multi-data joint adjoint. The package is built on Devito~\citep{Louboutin2019,Luporini2020} and uses the SciPy L-BFGS-B algorithm~\citep{Byrd1995} for the optimisation loop, with Dask driving the shot-parallel FWI runs. The source code, along with the FWI driver scripts required to reproduce every figure in this paper from public synthetic-data generators without any proprietary inputs, will be published upon acceptance.

\section*{Declaration of competing interest}

The authors declare that they have no known competing financial interests or personal relationships that could have appeared to influence the work reported in this paper.

\section*{Acknowledgements}
We would like to thank the sponsors of the Reservoir Characterization Project (RCP) at the Colorado School of Mines for the financial support of this work.

\bibliographystyle{IEEEtran}
\bibliography{refs}

\begin{thebibliography}{10}
\providecommand{\url}[1]{#1}
\csname url@samestyle\endcsname
\providecommand{\newblock}{\relax}
\providecommand{\bibinfo}[2]{#2}
\providecommand{\BIBentrySTDinterwordspacing}{\spaceskip=0pt\relax}
\providecommand{\BIBentryALTinterwordstretchfactor}{4}
\providecommand{\BIBentryALTinterwordspacing}{\spaceskip=\fontdimen2\font plus
\BIBentryALTinterwordstretchfactor\fontdimen3\font minus
  \fontdimen4\font\relax}
\providecommand{\BIBforeignlanguage}[2]{{%
\expandafter\ifx\csname l@#1\endcsname\relax
\typeout{** WARNING: IEEEtran.bst: No hyphenation pattern has been}%
\typeout{** loaded for the language `#1'. Using the pattern for}%
\typeout{** the default language instead.}%
\else
\language=\csname l@#1\endcsname
\fi
#2}}
\providecommand{\BIBdecl}{\relax}
\BIBdecl

\bibitem{Wang2018}
\BIBentryALTinterwordspacing
H.~F. Wang, X.~Zeng, D.~E. Miller, D.~Fratta, K.~L. Feigl, C.~H. Thurber, and
  R.~J. Mellors, ``Ground motion response to an {ML} 4.3 earthquake using
  co-located distributed acoustic sensing and seismometer arrays,''
  \emph{Geophysical Journal International}, vol. 213, no.~3, pp. 2020--2036, 06
  2018. [Online]. Available: \url{https://doi.org/10.1093/gji/ggy102}
\BIBentrySTDinterwordspacing

\bibitem{Mateeva2014}
\BIBentryALTinterwordspacing
A.~Mateeva, J.~Lopez, H.~Potters, J.~Mestayer, B.~Cox, D.~Kiyashchenko,
  P.~Wills, S.~Grandi, K.~Hornman, B.~Kuvshinov, W.~Berlang, Z.~Yang, and
  R.~Detomo, ``Distributed acoustic sensing for reservoir monitoring with
  vertical seismic profiling,'' \emph{Geophysical Prospecting}, vol.~62, no.~4,
  pp. 679--692, 2014. [Online]. Available:
  \url{https://onlinelibrary.wiley.com/doi/abs/10.1111/1365-2478.12116}
\BIBentrySTDinterwordspacing

\bibitem{Li2023}
\BIBentryALTinterwordspacing
J.~Li, W.~Zhu, E.~Biondi, and Z.~Zhan, ``Earthquake focal mechanisms with
  distributed acoustic sensing,'' \emph{Nature Communications}, vol.~14, p.
  4181, 2023. [Online]. Available:
  \url{https://doi.org/10.1038/s41467-023-39639-3}
\BIBentrySTDinterwordspacing

\bibitem{Liu2025}
\BIBentryALTinterwordspacing
F.~Liu, C.~Macesanu, H.~Xing, M.~Romanenko, G.~Zhan, C.~Calder\'on-Mac\'ias,
  and B.~Wang, ``Elastic full-waveform inversion: Enhance imaging for legacy
  and modern acquisition,'' \emph{The Leading Edge}, vol.~44, no.~5, pp.
  338--343, 2025. [Online]. Available:
  \url{https://doi.org/10.1190/tle44050338.1}
\BIBentrySTDinterwordspacing

\bibitem{Virieux2009}
\BIBentryALTinterwordspacing
J.~Virieux and S.~Operto, ``An overview of full-waveform inversion in
  exploration geophysics,'' in \emph{Geophysics Today: A Survey of the Field as
  the Journal Celebrates its 75th Anniversary}, S.~Fomel, Ed.\hskip 1em plus
  0.5em minus 0.4em\relax Society of Exploration Geophysicists, 2010, vol.~16,
  p.~0. [Online]. Available: \url{https://doi.org/10.1190/1.9781560802273}
\BIBentrySTDinterwordspacing

\bibitem{Pratt1998}
\BIBentryALTinterwordspacing
R.~G. Pratt, C.~Shin, and G.~J. Hick, ``Gauss--{Newton} and full {Newton}
  methods in frequency--space seismic waveform inversion,'' \emph{Geophysical
  Journal International}, vol. 133, no.~2, pp. 341--362, 05 1998. [Online].
  Available: \url{https://doi.org/10.1046/j.1365-246X.1998.00498.x}
\BIBentrySTDinterwordspacing

\bibitem{Brossier2009}
R.~Brossier, S.~Operto, and J.~Virieux, ``Seismic imaging of complex onshore
  structures by 2{D} elastic frequency-domain full-waveform inversion,''
  \emph{Geophysics}, vol.~74, no.~6, pp. WCC105--WCC118, 2009.

\bibitem{Operto2013}
\BIBentryALTinterwordspacing
S.~Operto, Y.~Gholami, V.~Prieux, A.~Ribodetti, R.~Brossier, L.~Metivier, and
  J.~Virieux, ``A guided tour of multiparameter full-waveform inversion with
  multicomponent data: From theory to practice,'' \emph{The Leading Edge},
  vol.~32, no.~9, pp. 1040--1054, 09 2013. [Online]. Available:
  \url{https://doi.org/10.1190/tle32091040.1}
\BIBentrySTDinterwordspacing

\bibitem{Tura2025}
\BIBentryALTinterwordspacing
A.~Tura, R.~Yermakov, O.~E. Aaker, {\O}.~Pedersen, A.~Damasceno, M.~R. Braga,
  S.~L.~S. Hoey, L.~Ambati, N.~N.~A. Rahman, S.~K. Chandola, A.~R. Ghazali,
  M.~F.~A. Halim, and B.~Olofsson, ``Modern case study demonstrations of the
  industry value of {PS}-wave data,'' \emph{First Break}, vol.~43, no.~11, pp.
  75--84, 2025. [Online]. Available:
  \url{https://www.earthdoc.org/content/journals/10.3997/1365-2397.fb2025089}
\BIBentrySTDinterwordspacing

\bibitem{Kohn2014}
\BIBentryALTinterwordspacing
D.~K{\"o}hn, D.~De~Nil, A.~Kurzmann, A.~Przebindowska, and T.~Bohlen, ``On the
  influence of model parametrization in elastic full waveform tomography,''
  \emph{Geophysical Journal International}, vol. 191, no.~1, pp. 325--345, 10
  2012. [Online]. Available:
  \url{https://doi.org/10.1111/j.1365-246X.2012.05633.x}
\BIBentrySTDinterwordspacing

\bibitem{Modrak2016}
\BIBentryALTinterwordspacing
R.~Modrak and J.~Tromp, ``Seismic waveform inversion best practices: regional,
  global and exploration test cases,'' \emph{Geophysical Journal
  International}, vol. 206, no.~3, pp. 1864--1889, 09 2016. [Online].
  Available: \url{https://doi.org/10.1093/gji/ggw202}
\BIBentrySTDinterwordspacing

\bibitem{Pan2019}
\BIBentryALTinterwordspacing
W.~Pan, K.~A. Innanen, and Y.~Geng, ``Elastic full-waveform inversion and
  parametrization analysis applied to walk-away vertical seismic profile data
  for unconventional (heavy oil) reservoir characterization,''
  \emph{Geophysical Journal International}, vol. 213, no.~3, pp. 1934--1968, 06
  2018. [Online]. Available: \url{https://doi.org/10.1093/gji/ggy087}
\BIBentrySTDinterwordspacing

\bibitem{Eaid2020}
\BIBentryALTinterwordspacing
M.~V. Eaid, S.~D. Keating, and K.~A. Innanen, ``Multiparameter seismic elastic
  full-waveform inversion with combined geophone and shaped fiber-optic cable
  data,'' \emph{Geophysics}, vol.~85, no.~6, pp. R537--R552, 11 2020. [Online].
  Available: \url{https://doi.org/10.1190/geo2020-0170.1}
\BIBentrySTDinterwordspacing

\bibitem{Lellouch2020}
A.~Lellouch and B.~L. Biondi, ``Seismic applications of downhole {DAS},''
  \emph{Sensors (Basel)}, vol.~21, no.~9, p. 2897, 04 2021, pMID: 33919095;
  PMCID: PMC8122346.

\bibitem{Sayed2020}
A.~Sayed, S.~Ali, and R.~R. Stewart, ``Distributed acoustic sensing ({DAS}) to
  velocity transform and its benefits,'' in \emph{SEG Technical Program
  Expanded Abstracts 2020}.\hskip 1em plus 0.5em minus 0.4em\relax Society of
  Exploration Geophysicists, 2020, pp. 3788--3792.

\bibitem{Zhou2024}
\BIBentryALTinterwordspacing
W.~Zhou, ``Direct full waveform inversion of {DAS} fiber-optic data,'' vol.
  2024, no.~1, pp. 1--5, 2024. [Online]. Available:
  \url{https://www.earthdoc.org/content/papers/10.3997/2214-4609.202410385}
\BIBentrySTDinterwordspacing

\bibitem{Virieux1986}
\BIBentryALTinterwordspacing
J.~Virieux, ``{P-SV} wave propagation in heterogeneous media; velocity-stress
  finite-difference method,'' \emph{Geophysics}, vol.~51, no.~4, pp. 889--901,
  04 1986. [Online]. Available: \url{https://doi.org/10.1190/1.1442147}
\BIBentrySTDinterwordspacing

\bibitem{Komatitsch2007}
\BIBentryALTinterwordspacing
D.~Komatitsch and R.~Martin, ``An unsplit convolutional perfectly matched layer
  improved at grazing incidence for the seismic wave equation,''
  \emph{Geophysics}, vol.~72, no.~5, pp. SM155--SM167, 08 2007. [Online].
  Available: \url{https://doi.org/10.1190/1.2757586}
\BIBentrySTDinterwordspacing

\bibitem{Hartog2017}
\BIBentryALTinterwordspacing
A.~H. Hartog, \emph{An Introduction to Distributed Optical Fibre Sensors},
  1st~ed.\hskip 1em plus 0.5em minus 0.4em\relax Boca Raton: CRC Press, 05
  2017. [Online]. Available: \url{https://doi.org/10.1201/9781315119014}
\BIBentrySTDinterwordspacing

\bibitem{Louboutin2019}
\BIBentryALTinterwordspacing
M.~Louboutin, M.~Lange, F.~Luporini, N.~Kukreja, P.~A. Witte, F.~J. Herrmann,
  P.~Velesko, and G.~J. Gorman, ``{Devito} (v3.1.0): an embedded
  domain-specific language for finite differences and geophysical
  exploration,'' \emph{Geoscientific Model Development}, vol.~12, no.~3, pp.
  1165--1187, 2019. [Online]. Available:
  \url{https://gmd.copernicus.org/articles/12/1165/2019/}
\BIBentrySTDinterwordspacing

\bibitem{Luporini2020}
\BIBentryALTinterwordspacing
F.~Luporini, M.~Lange, M.~Louboutin, N.~Kukreja, J.~H{\"u}ckelheim, C.~Yount,
  P.~Witte, P.~H.~J. Kelly, F.~J. Herrmann, and G.~J. Gorman, ``Architecture
  and performance of {Devito}, a system for automated stencil computation,''
  2020. [Online]. Available: \url{https://arxiv.org/abs/1807.03032}
\BIBentrySTDinterwordspacing

\bibitem{Holberg1987}
\BIBentryALTinterwordspacing
O.~Holberg, ``Computational aspects of the choice of operator and sampling
  interval for numerical differentiation in large-scale simulation of wave
  phenomena,'' \emph{Geophysical Prospecting}, vol.~35, no.~6, pp. 629--655,
  1987. [Online]. Available:
  \url{https://onlinelibrary.wiley.com/doi/abs/10.1111/j.1365-2478.1987.tb00841.x}
\BIBentrySTDinterwordspacing

\bibitem{Tarantola1984}
\BIBentryALTinterwordspacing
A.~Tarantola, ``Inversion of seismic reflection data in the acoustic
  approximation,'' \emph{Geophysics}, vol.~49, no.~8, pp. 1259--1266, 08 1984.
  [Online]. Available: \url{https://doi.org/10.1190/1.1441754}
\BIBentrySTDinterwordspacing

\bibitem{Plessix2006}
\BIBentryALTinterwordspacing
R.-E. Plessix, ``A review of the adjoint-state method for computing the
  gradient of a functional with geophysical applications,'' \emph{Geophysical
  Journal International}, vol. 167, no.~2, pp. 495--503, 11 2006. [Online].
  Available: \url{https://doi.org/10.1111/j.1365-246X.2006.02978.x}
\BIBentrySTDinterwordspacing

\bibitem{Tarantola1986}
\BIBentryALTinterwordspacing
A.~Tarantola, ``A strategy for nonlinear elastic inversion of seismic
  reflection data,'' \emph{Geophysics}, vol.~51, no.~10, pp. 1893--1903, 10
  1986. [Online]. Available: \url{https://doi.org/10.1190/1.1442046}
\BIBentrySTDinterwordspacing

\bibitem{Bunks1995}
\BIBentryALTinterwordspacing
C.~Bunks, F.~M. Saleck, S.~Zaleski, and G.~Chavent, ``Multiscale seismic
  waveform inversion,'' \emph{Geophysics}, vol.~60, no.~5, pp. 1457--1473, 10
  1995. [Online]. Available: \url{https://doi.org/10.1190/1.1443880}
\BIBentrySTDinterwordspacing

\bibitem{Byrd1995}
\BIBentryALTinterwordspacing
R.~H. Byrd, P.~Lu, J.~Nocedal, and C.~Zhu, ``A limited memory algorithm for
  bound constrained optimization,'' \emph{SIAM Journal on Scientific
  Computing}, vol.~16, no.~5, pp. 1190--1208, 1995. [Online]. Available:
  \url{https://doi.org/10.1137/0916069}
\BIBentrySTDinterwordspacing

\bibitem{Plessix2004}
R.-E. Plessix and W.~A. Mulder, ``Frequency-domain finite-difference
  amplitude-preserving migration,'' \emph{Geophysical Journal International},
  vol. 157, no.~3, pp. 975--987, 2004.

\bibitem{Dask2016}
{Dask Development Team}, ``Dask: Library for dynamic task scheduling,''
  \url{http://dask.pydata.org}, 2016.

\bibitem{10.1007/10968987_3}
A.~B. Yoo, M.~A. Jette, and M.~Grondona, ``Slurm: Simple linux utility for
  resource management,'' in \emph{Job Scheduling Strategies for Parallel
  Processing}, D.~Feitelson, L.~Rudolph, and U.~Schwiegelshohn, Eds.\hskip 1em
  plus 0.5em minus 0.4em\relax Berlin, Heidelberg: Springer Berlin Heidelberg,
  2003, pp. 44--60.

\bibitem{Wang2004}
Z.~Wang, A.~C. Bovik, H.~R. Sheikh, and E.~P. Simoncelli, ``Image quality
  assessment: from error visibility to structural similarity,'' \emph{IEEE
  Transactions on Image Processing}, vol.~13, no.~4, pp. 600--612, 04 2004,
  pMID: 15376593.

\bibitem{Martin2006}
\BIBentryALTinterwordspacing
G.~S. Martin, R.~Wiley, and K.~J. Marfurt, ``{Marmousi2}: An elastic upgrade
  for {Marmousi},'' \emph{The Leading Edge}, vol.~25, no.~2, pp. 156--166, 02
  2006. [Online]. Available: \url{https://doi.org/10.1190/1.2172306}
\BIBentrySTDinterwordspacing

\end{thebibliography}

\end{document}